\documentclass{aa}  

\usepackage{graphicx} 
\usepackage{longtable}
\usepackage{lscape}
\usepackage{pdfsync}
\usepackage{bbding}
\usepackage[export]{adjustbox}
\usepackage{natbib} 
\usepackage{txfonts}
\begin{document}
   \title{Active galactic nuclei synapses:}

   \subtitle{X-ray versus optical classifications using artificial neural networks}

   \author{O. Gonz\'alez-Mart\'in\inst{1,2}\thanks{Juan de la Cierva Fellow (\email{omairagm@iac.es})}
   \and
   D. D\'iaz-Gonz\'alez\inst{3}
   \and
   J.~A. Acosta-Pulido\inst{1,2}
\and 
   J. Masegosa\inst{4} 
    \and  \\ 
   I.E. Papadakis\inst{5,6}
     \and
   J.M. Rodr\'iguez-Espinosa\inst{1,2}
 \and
  I. M\'arquez\inst{4}
   \and
    L. Hern\'andez-Garc\'ia\inst{4}
          }

   \institute{
Instituto de Astrof\'isica de Canarias (IAC), C/V\'ia L\'actea, s/n, E-38205 La Laguna, Spain. \and 
Departamento de Astrof\'isica, Universidad de La Laguna (ULL), E-38205 La Laguna, Spain. \and 
Shidix Technologies, E-38320, La Laguna, Spain. \and
Instituto de Astrof\'isica de Andaluc\'ia, CSIC, C/ Glorieta de la Astronom\'ia, s/n E-18005 Granada, Spain. \and
Physics Department, University of Crete, PO Box 2208, 710 03 Heraklion, Crete, Greece. \and
IESL, Foundation for Research and Technology, 711 10, Heraklion, Crete, Greece.
             }

   \date{Accepted 3rd April 2014}

 
  \abstract
   { Many classes of active galactic nuclei (AGN) have been defined entirely throughout optical wavelengths while the X-ray spectra have been very useful to investigate their inner regions. However, optical and X-ray results show many discrepancies that have not been fully understood yet. }
   {The main purpose of the present paper is to study the ``synapses'' (i.e., connections) between the X-ray and optical AGN classifications. }
   {For the first time, the newly implemented {\sc efluxer} task allowed us to analyse broad band X-ray spectra of a sample of emission line nuclei without any prior spectral fitting. Our sample comprises 162 spectra observed with \emph{XMM}-Newton/pn of 90 local emission line nuclei in the Palomar sample. It includes, from the optical point of view, starbursts (SB), transition objects (T2), low ionisation nuclear emission line regions (L1.8 and L2), and Seyfert nuclei (S1, S1.8, and S2). We use artificial neural networks (ANNs) to study the connection between X-ray spectra and the optical classes. }
   {Among the training classes, the ANNs are 90\% efficient at classifying the S1, S1.8, and SB classes. The S1 and S1.8 classes show a negligible SB-like component contribution with a wide range of those from S1- and S1.8-like components. We suggest that this broad range of values is related to a large degree of obscuration in the X-ray regime. When including all the objects in our sample, the S1, S1.8, S2, L1.8, L2/T2/SB-AGN (SB with indications of AGN activity in the literature), and SB classes have similar average X.ray spectra, but these average spectra can be distinguished from class to class. The S2 (L1.8) class is linked to the S1.8 (S1) class with larger SB-like component than the S1.8 (S1) class. The L2, T2, and SB-AGN classes conform a class in the X-rays similar to the S2 class albeit with larger fractions of SB-like component. We argue that this SB-like component might come from the contribution of the host galaxy emission to the X-rays, which is large when the AGN is weak. Up to 80\% of the emission line nuclei and, on average, all the optical classes included in our sample show a non-negligible fraction of S1-like or S1.8-like component. Thus, an AGN-like component seems to be present in the vast majority of the emission line nuclei in our sample. }
   {The ANN trained in this paper is not only useful to study the synergies between the optical and X-ray classifications, but also could be used to infer optical properties from X-ray spectra in surveys like \emph{eRosita}.}


   \maketitle
%

\section{Introduction}\label{sec:introduction}

At optical wavelengths emission line galaxies can be grouped into HII nuclei, active galactic nuclei (AGN), galaxies with low-ionisation nuclear emission line regions (LINERs), and transition objects \citep[whose optical spectra are intermediate between those of pure LINERs and HII regions; see][for a review]{Ho08}. Optical spectroscopic studies have shown that only 10\% of nearby galaxies are Seyferts, while LINERs and transition objects account to no more than 20\% and 10\% of them, respectively \citep[e.g., Palomar Survey by][]{Ho97}.

HII nuclei are powered by a compact star forming region. In AGN, the major energy source is assumed to be accretion of matter into a super-massive black hole (SMBH). The nature of the main energy source in LINERs (and transition objects) is not settled yet. They might be low-luminosity AGN (LLAGN), in which case, they will constitute the main fraction of the AGN population \citep{Heckman80,Ho97}. However, other emission mechanisms like shock heating \citep{Dopita95}, OB stars in compact nuclear star clusters \citep{Terlevich85}, or pre-main sequence stars ionisation \citep{Cid-Fernandes04} have also been proposed.

AGN are traditionally divided into two main classes, namely Type-1 and Type-2 objects, based on the existence (Type-1) or not (Type-2) of broad permitted lines (FWHM$\rm{>}$2000 km $\rm{s^{-1}}$). The so-called unification model (UM) proposes that both types of AGN are essentially the same objects viewed at different angles \citep{Antonucci93,Urry95}. An optically thick dusty torus surrounding the central source would be responsible for blocking the region where these broad emission lines are produced (the broad line region, BLR) in Type-2 Seyferts. Therefore, Type-2 Seyferts are essentially Type-1 Seyferts blocked by the dusty torus along the line of sight (LOS) to the observer. A strong observational evidence in favour of a unification between Type-1 and Type-2 Seyferts was the discovery of broad optical lines in the polarised spectrum of the archetypal Type-2 Seyfert, NGC\,1068 \citep{Antonucci85}. The torus must not be spherically symmetric, in order to obscure the BLR, allowing at the same time the region producing the permitted narrow lines (known as narrow-line region, NLR) to reach us from the same LOS. The locus of this obscuring material was initially postulated at parsec scales and confirmed by modelling the spectral energy distribution (SED) of Seyferts \citep[e.g.,][]{Ramos-Almeida11,Alonso-Herrero11} and by interferometric observations \citep[e.g., Circinus galaxy,][]{Tristram07}. Such scales are unreachable with the current instrumentation, so the torus morphology can only be inferred by indirect measurements.

Although the UM is widely accepted for many classes of Seyferts, there is still no consensus on its general applicability for all members of each class \citep[see][for a review]{Bianchi12}. An example of this mismatch is the so-called `optically elusive' AGN \citep{Maiolino98}. These elusive AGN are nuclear hard X-ray sources whose intrinsic luminosities are in the Seyfert range but they lack of optical Seyfert-like signatures. Another example is that about half of the brightest Type-2 Seyferts are characterised by the lack of BLR even with high-quality spectro-polarimetric data \citep[known as `True Type-2' Seyferts,][]{Tran01,Tran03}. These Type-2 Seyferts without BLR are expected to occur theoretically at low accretion rates or low luminosities \citep{Elitzur09}. 

With respect to LINERs, even if they are powered predominately by accretion into a SMBH, it is unclear whether the UM can also apply to these LLAGN. Indeed, both a different accretion mode and large amounts of obscuration have been proposed to explain the differences between LINERs and Seyferts \citep{Gonzalez-Martin06,Gonzalez-Martin09a,Gonzalez-Martin09b,Younes11,Hernandez-Garcia13}.

X-rays in AGN are thought to originate in the innermost region of the accretion flow and are also thought to be affected by the obscuring material along the LOS. X-ray observations of AGN have provided additional evidence in favour of the UM. For example, the obscuring material along the LOS (measured at X-rays by the hydrogen column density, $\rm{N_H}$) is substantially larger in Type-2 Seyferts than in Type-1 Seyferts \citep[e.g.,][]{Maiolino98,Risaliti99,Panessa06,Cappi06}. Although modelling of X-ray spectra is one of the best ways to estimate the obscuration, it has also some caveats. For example, the obscuration measured in Seyferts depends on the model used for the underlying X-ray continuum.

The main aim of this paper is to investigate if objects in different (optical) classes have similar X-ray spectra, and if they do, whether their average X-ray spectrum differs between the different classes or not. Furthermore, we compare the average X-ray spectra of these classes in a model independent way. Consequently, instead of fitting each individual spectrum with a suitable model, we chose to use artificial neural networks (ANNs). 

We have selected for our analysis the X-ray spectra of 90 well-classified emission line nuclei included in the optically classified sample of nearby galaxies presented by \citet{Ho97}. We used ANNs to classify their X-ray spectrum and compare the average spectra of each class, without any model pre-assumptions. The main questions we address in this paper are the following: (1) how do optical classes ``behave'' at X-rays? in other words, do objects of the same (optical) class have the same X-ray spectrum (on average), and if yes, are the average X-ray spectra of the various optical classes the same or not? (2) If they are different, can we understand what is the main physical parameter that drives those differences? and (3) are AGN-like nuclei present in all emission line nuclei in nearby galaxies? does this include those galaxies that have absent or weak AGN signatures at optical wavelengths?

Section \ref{sec:sample} gives the details on the selected sample and Section \ref{sec:reduction} the technical details of the reduction process. In Section \ref{sec:ANN} we describe the methodology and the main results of the ANN are presented in Section \ref{sec:ANNresults}. These results are discussed in Section \ref{sec:discussion} and summarised in Section \ref{sec:summary}. Along the paper a value of $\rm{H_{0} = 75}$ km s$^{-1}$ Mpc$^{-1}$ is assumed. 

\section{Sample}\label{sec:sample}

We have used the Palomar sample, a catalog of optical nuclear spectra reported by \citet{Ho97}. This is the largest sample of galaxy nuclei with optical spectra homogeneously observed in the nearby Universe up to date. They presented measurements of the spectroscopic parameters for 418 emission-line nuclei. The sample contains most of the bright galaxies ($\rm{M_{B} < 12}$) in the nearby Universe. Since our work will be based on the optical classification of AGN, we consider the homogeneous analysis performed by \citet{Ho97} as ideal for our purpose. 

We have obtained all the available (up to December 2012) \emph{XMM}-Newton\footnote{We have used the HEASARC archive to download the data at http://heasarc.nasa.gov} data for the objects in the Palomar sample. We initially included 436 observations in our sample. We excluded the observations where the source of interest for our analysis was out of the field of view, not detected, or close to the gap between chips in the EPIC-pn detector. We then excluded the observations for which the pileup\footnote{Pileup occurs on X-ray CCDs when several photons hit the detector at the same place between two read-outs \citep{Ballet99}. } was higher than 5\% (NGC\,1275, ObsID 0305780101 and NGC\,4486, ObsID 0200920101). We only considered spectra with more than $\sim$500 net counts in the 0.5--10~keV band. We imposed this restriction to include only high S/N data.

Our final sample contains 162 observations for 90 emission line nuclei. This represents $\sim$20\% of the sample published by \citet{Ho97}. Table \ref{tab:sample} shows the observational details of the X-ray data of the sample: object name (Col.~2), identifier of the observation -- ObsID (Col~.3), optical class (Col.~4), net exposure time (Col.~5), and net number counts (Col.~6). The optical classification is that reported in \citet{Ho97}. 

Our sample includes ten S1 objects (optically classified as `S1', `S1.2', and `S1.5'), eight S1.8 objects  (optically classified as `S1.8', and `S1.9'), nine S2 sources (optically classified as `S2', `S2:', and `S2::'), 11 L1.8 objects (optically classified as `L1.9'), 17 L2 objects (optically classified as `L2', `L2:',`L2::', and `S2/L'), 11 T2 objects (optically classified as `T2', `T2:', and `T2/S'), and 24 SB objects (optically classified as `H' and  `H:'). The optical classes were classified by \citet{Ho97} using BPT diagrams \citep[named after ``Baldwin, Phillips \& Telervich'',][]{Baldwin81}. These diagrams are based on nebular emission line ratios used to distinguish the ionisation mechanism of the ionising gas. The better known version consists on a combination of three diagrams: [NII]$\rm{\lambda}$6584/H$\rm{\alpha}$ versus [OIII]$\rm{\lambda}$5007/H$\rm{\beta}$, [SII]$\rm{\lambda}$6717,6731/H$\rm{\alpha}$ versus [OIII]$\rm{\lambda}$5007/H$\rm{\beta}$ and [OI]$\rm{\lambda}$6300/H$\rm{\alpha}$ versus [OIII]$\rm{\lambda}$5007/H$\rm{\beta}$. The classifications into Type 1, 1.2, 1.5, 1.8, and 1.9 were made based on the presence and strength of broad components for $\rm{H\alpha}$ and $\rm{H\beta}$ lines. Note that here the L1.8 class refer, for consistency with the S1.8 class, to objects belonging to the L1.8 and L1.9 optical type; however the L1.8 sample is actually made only of objects optically classified as L1.9.

AGN signatures (mostly from X-ray spectral studies) have been discovered in half of the SB objects in our sample (12 out of 24), after their classification by \citet{Ho97}. Six S2 nuclei belong to the category of True Type-2 Seyferts. Furthermore, eight objects (classified as S1, S1.8, S2, L1.8, L2, or SB) show a hydrogen column density in the Compton thick regime (i.e. $\rm{N_{H}>1.5\times 10^{24}cm^{-2}}$). This information, together with the corresponding references, is included in Col. 11 of Table \ref{tab:sample}.

\section{X-ray data processing}\label{sec:reduction}

The \emph{XMM-Newton} data were reduced with the latest SAS version (v12.0.1), using the most up to date calibration files available. In this paper, only EPIC/pn \citep{Struder01} data have been analysed because of their higher count rate and lower distortion due to pileup.

Time intervals of quiescent particle background were screened from the net source spectrum by excluding time intervals above 3$\rm{\sigma}$ of the median value for the background light curve. The nuclear positions were retrieved from NED and source counts in each case were accumulated from a circular region of radii between 15${{\hbox{$^{\prime\prime}$}}-50{\hbox{$^{\prime\prime}$}}}$ (300-1000 pixels). These radii were chosen to avoid nearby sources and to sample most of the PSF according to the observing mode. The background region was selected using a source-free circular region on the same CCD chip than the source with an automatic routine created with IDL. We selected only single and double pixel events (i.e., patterns of 0-4). Bad pixels and events too close to the edges of the CCD chip were rejected using ``FLAG=0''. The regions were extracted with the SAS {\sc evselect} task. pn redistribution matrix and effective areas were calculated with {\sc rmfgen} and {\sc arfgen} tasks, respectively.

Pileup affects both flux measurements and spectral characterisation of bright sources \citep{Ballet01}. The pileup has been estimated with the {\sc pimms} software using the 0.5-10 keV flux interval and assuming a power-law model with slope $\Gamma=2.1$ (canonical value for AGN) and the setting of each observation. Note that observations with pileup fractions larger than 5\% were previously excluded from our sample (see Section \ref{sec:sample}). Only two observations showed a pile up fraction below 5\%: NGC\,1275 (ObsID 0085110101) and NGC\,4486 (ObsID 0114120101) with 3.2\% and 2.2\% pileup, respectively. Thus, the pileup is negligible in our sample.

The spectra were flux-calibrated using the {\sc efluxer} task within the SAS. The final spectral range goes between 0.5--10.0 keV with energy bins of $\rm{\Delta E =}$0.05 keV. Note that we excluded data below 0.5 keV since {\sc efluxer} seems to be less accurate at such energies. These final spectra are expressed in luminosity units (erg/s) and redshifted to rest-frame according to the distance of the source (see Table \ref{tab:sample}). The flux-calibrated spectra for the entire sample are provided in the Appendix B.

\section{Artificial neural network}\label{sec:ANN}

As we explained in Section \ref{sec:introduction}, we did not follow the standard procedure of fitting the X-ray spectra with a model in order to avoid the possibility that the results may be affected by model-dependent degeneracies. Instead we chose to use ANNs. Briefly, ANNs are computing algorithms resembling to some extent the behaviour of the brain. They consist on processing units, \emph{neurones}, with multiple signal transmitter connections, organised as a network. These connections have adaptable strengths, \emph{synaptic weights}, which modify the signal transmitted to (and from) each neurone. The training of the network is the process of adjusting weights, so that the network learns how to solve a specific problem. We describe this process in the following subsections. 

The code used to implement the ANN is the Python-Based Reinforcement Learning, Artificial Intelligence and Neural (PyBrain) network library \citep{Schaul10}. PyBrain\footnote{http://pybrain.org/pages/home} is a modular Machine Learning Library for Python. 

\subsection{Inputs, outputs, and the network training}\label{sec:inputs}\label{sec:outputs}

The primary inputs for this study are the X-ray spectra of the objects in our sample. These spectra have been extracted using standard X-ray procedures as explained in Section \ref{sec:reduction} and then converted to physical units with the algorithm {\sc efluxer} within the SAS.

The training process is set to classify the X-ray spectra of the sources within the SB and S1 optical classes. We chose these classes for the training of the network, as the objects belonging to them are supposed to be representative of objects where accretion (Seyferts) and star-forming related processes (SBs) are the main source of power, respectively. To study the connection in X-rays between Type-1 and Type-2 optical classes, one should ideally use S1 and S2 samples. However, the S2 class is made by several types of objects whose nature might be controversial. Some objects are heavily obscured (with negligible emission in X-rays) while others may lack the BLR \citep[see Table \ref{tab:sample} in this paper and][for a review]{Bianchi12}. We therefore choose to use the S1.8 sample as a third training set since this represents a more homogeneous class in X-rays than the S2 class. We therefore used the following three classes for our training sets: 

\vspace{0.1cm}

$\bullet$ \underline{`S1-Training'}: We ascribed to this class all the objects within the S1 class. This training set includes nine AGN. We have excluded NGC\,1275 because of the strong contribution of the diffuse emission from the centre of the galaxy cluster \citep[see][]{Sanders05}. 

\vspace{0.1cm}

$\bullet$ \underline{`S1.8-Training'}: We ascribed to this class all the objects in the S1.8 class. This training set contains seven AGN. We excluded NGC\,1068 because it is a Compton-thick source and, therefore, the primary AGN emission is not seen at the energy range analysed in this study. 

\vspace{0.1cm}

$\bullet$ \underline{`SB-Training'}:  Only the SB class (objects marked with `H' in Table \ref{tab:sample}) is included in this training set, avoiding the objects classified as SB-AGN (see Sect. \ref{sec:sample}). We excluded IC\,10 because this galaxy hosts a ULX included in the PSF of \emph{XMM}-Newton. This training set contains 11 SBs.

\vspace{0.1cm}

For objects with more than one observations, we chose that with the largest luminosity. We tested that the selection of another observation of the same object does not change the final classification substantially. Thus, the training set contains a total of 27 spectra (one per object). All the observations used for the training process are marked as `TR' in Col.~7 of Table~\ref{tab:sample}. 

The optical classification is recorded in the network outputs using a vector of three elements $\rm{\nu \equiv [\nu_{S1}, \nu_{S1.8}, \nu_{SB}]}$. During the training process, we used the vectors $\rm{\nu= [100,0,0]}$, $\rm{\nu= [0,100,0]}$, and $\rm{\nu= [0,0,100]}$ to define the S1-, S1.8- and SB-Training groups, respectively. 

The training method used is the supervised regression training (`SupervisedDataSet' within PyBrain) with one hidden layer. In this method the training process is carried out until the network reliably matches the `a priori' known optical classification. 

\subsection{The ANN classification for the full data set}\label{sec:training}

The ANN training was able to converge to a solution. We then classified all the available spectra in our sample (including those used for the training process). 

For each spectrum the ANN gave a set of three elements. Each one of these elements can be considered as an indicator of the resemblance of an X-ray spectrum to the trained X-ray spectra of the S1, S1.8, and SB classes. For example, a spectrum fully consistent with the S1, S1.8, or SB classes should show a vector equal to (100,0,0), (0,100,0), or (0,0,100), respectively. If on the other hand, a spectrum is the combination of the S1, S1.8 and SB-Training sets, we would expect that the sum of $\rm{\nu_{S1}}$, $\rm{\nu_{S1.8}}$, and $\rm{\nu_{SB}}$ is equal to 100 (or consistent within errors). The larger is the number of $\rm{\nu_{S1}}$, $\rm{\nu_{S1.8}}$, or $\rm{\nu_{SB}}$ the closer the spectrum will resemble to the X-ray spectra of the S1- S1.8- or SB-Training sets, respectively. 

We also assigned errors ($\rm{\Delta \nu}$) to each of these three elements of the ANN, for each spectrum, using Monte Carlo simulations. We trained-and-classified the objects 100 times converging to individual solutions. For each training we obtained 1000 solutions randomly varying the spectra within the measurement error bars for each energy bin. The final solution is the mean value for the 100 thousand runs (i.e., 100 times 1000 solutions) and $\rm{\Delta \nu}$ is its standard deviation. Cols. 8, 9, and 10 in Table \ref{tab:sample} show the results for $\rm{\nu_{S1}}$, $\rm{\nu_{S1.8}}$, and $\rm{\nu_{SB}}$, respectively. 

Values significantly above 100 or below 0 will imply that the spectra cannot be reproduced with the training classes. None of the objects in our class show ANN components above 100 or below 0 at $\rm{\sim1.5\sigma}$ level). Thus, all of them can be characterised by a combination of the training sets. 

The efficiency of the network on the training process can be estimated by its success on classifying the training sets. Indeed, it has successfully classified 25 out of the 27 spectra within 10\% error (typical error obtained by the ANN). Thus, the efficiency of the network is $\rm{\sim}$90\%. Only one S1 (NGC\,4639) and two S1.8s (NGC\,4168 and NGC\,4565) were misclassified showing $\rm{\nu_{SB}>10}$. However, they show $\rm{\nu_{S1}}$ and $\rm{\nu_{S1.8}}$ fully consistent with their training sets within the errors (i.e., S1-Training for NGC\,4639 and S1.8-Training for NGC\,4168 and NGC\,4565). 

\begin{figure}
\centering 
\includegraphics[width=1.\columnwidth]{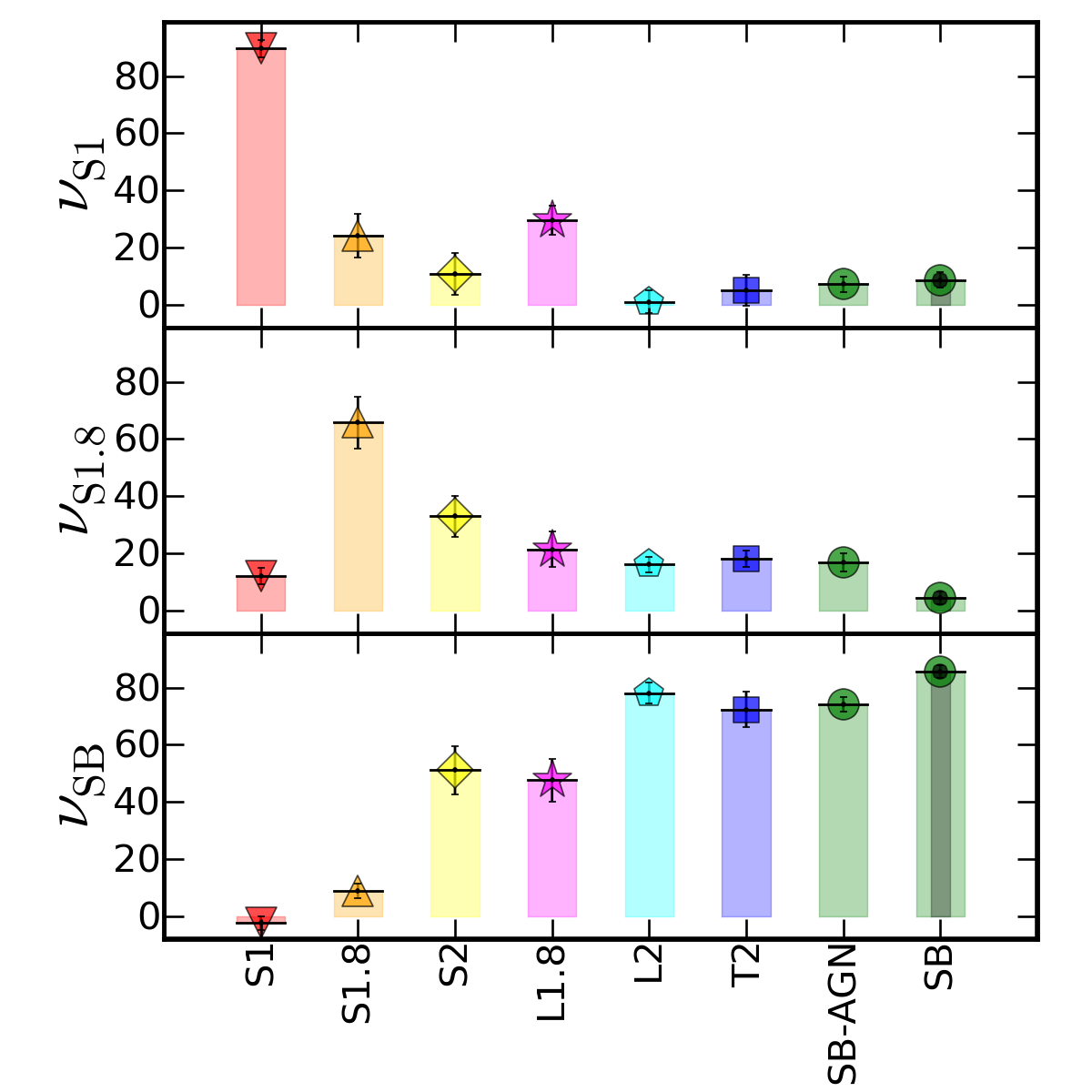}
\caption{Histogram of the mean value of the ANN components for each optical class. Error bars represent one sigma over the mean for each distribution. The optical classes are shown as: S1 (red up-side down triangles), S1.8 (orange triangles), S2 (yellow diamonds), L1.8 (purple stars), L2 (light blue pentagons), T2 (dark blue squares), SB-AGN (green circles), and SB (green circles with small black dots).}
\label{fig:ANNhist}
\end{figure}

\section{Results}\label{sec:ANNresults}

\begin{table*}
\centering
\begin{tabular}{l c c c c c c c c}
\hline \hline
        & \multicolumn{2}{c}{$\rm{\nu_{S1}}$} & & \multicolumn{2}{c}{$\rm{\nu_{S1.8}}$} & &  \multicolumn{2}{c}{$\rm{\nu_{SB}}$} \\ \cline{2-3} \cline{5-6} \cline{8-9}
        &     Mean            & Median &   &  Mean            & Median &  &   Mean            & Median \\ \hline
      S1 &   89.7 $\pm$    3.0 &   98.0 & &  12.1 $\pm$    2.9 &    3.3 & &  -2.3 $\pm$    2.4 &   -3.1 \\ 
   S1.8 &   24.2 $\pm$    7.7 &   10.5 &  & 65.8 $\pm$    9.0 &   78.7 &  &  9.0 $\pm$    2.6 &    6.4 \\ 
     S2 &   10.8 $\pm$    7.4 &    3.9 &  & 33.0 $\pm$    7.3 &   36.8 &  & 51.1 $\pm$    8.5 &   49.9 \\ 
   L1.8 &   29.6 $\pm$    5.2 &   33.2 &  & 21.4 $\pm$    6.2 &   10.5 &  & 47.6 $\pm$    7.4 &   49.0 \\ 
     L2 &    1.0 $\pm$    3.9 &    4.3 &  & 16.1 $\pm$    2.7 &   15.2 &  & 78.1 $\pm$    3.6 &   79.5 \\ 
     T2 &    5.0 $\pm$    5.5 &    8.7 &  & 18.1 $\pm$    2.7 &   19.4 & &  72.4 $\pm$    6.3 &   66.5 \\ 
 SB-AGN &    7.1 $\pm$    2.8 &    9.7 & &  16.9 $\pm$    3.2 &   17.0 & &  74.2 $\pm$    2.5 &   75.1 \\ 
 SB &    8.7 $\pm$    2.7 &    6.0 &  &  4.4 $\pm$    2.3 &    0.7 &  & 85.7 $\pm$    2.3 &   89.5 \\ \hline
\end{tabular}
\setcounter{table}{1}
\caption{Mean and median values for the ANN components per optical class.}  
\label{tab:meanANN}
\end{table*}

\subsection{Mean value of the ANN components per optical class}

First we present our results regarding the average value of the ANN ($\rm{\overline{\nu}_{S1}}$, $\rm{\overline{\nu}_{S1.8}}$, and $\rm{\overline{\nu}_{SB}}$) for each optical class. The mean, its error, and the median values for the ANN components per optical class are shown in Table \ref{tab:meanANN}. Fig.~\ref{fig:ANNhist} shows these mean values (and the errors as error bars) as a function of optical classes. The error of the mean of each ANN component is very small in all the optical classes. This implies that all the sources in each class have similar X-ray spectra. Secondly, the mean values are not the same in all classes. Therefore, the mean X-ray spectrum is not the same for all of them. We also obtain that: 

\begin{itemize}
\item Among the training classes, both S1 and S1.8 classes show low $\rm{\overline{\nu}_{SB}}$. The S1 class shows large $\rm{\overline{\nu}_{S1}}$ and low $\rm{\overline{\nu}_{S1.8}}$; the opposite is true for the S1.8 class. Similarly, the SB class shows large $\rm{\overline{\nu}_{SB}}$ and low $\rm{\overline{\nu}_{S1}}$ and $\rm{\overline{\nu}_{S1.8}}$. This was expected as we have trained the network to achieve this objective. However, we used all the spectra and not only those used for the training process. Thus, it seems that any eventual flux variations of S1, S1.8, and SB are not associated with spectral variations that can alter dramatically the shape of their X-ray spectra.
\item The S2, L1.8, L2, T2, and SB-AGN classes are not compatible with any of the trained classes (i.e., S1, S1.8, or SB). In fact, they can be interpreted as a combination of two out of the three ANN components (see Section \ref{sec:ANNcorr}).
\item The S2 class is not consistent with either the S1 or the S1.8 classes. On average S2 objects show very low $\rm{\overline{\nu}_{S1}}$, a $\rm{\overline{\nu}_{S1.8}}$ which lies between that of the S1 and the S1.8 classes, and $\rm{\overline{\nu}_{SB}}$ significantly larger than the respective mean value for the S1 and S1.8 classes (see Table \ref{tab:meanANN}). 
\item The L1.8 and L2 classes are not similar. In fact, their $\rm{\overline{\nu}_{S1.8}}$ and $\rm{\overline{\nu}_{SB}}$ values are similar to those of the S2 class. As a class though, L1.8 can be distinguished from S2, because their average $\rm{\overline{\nu}_{S1}}$ is larger ($\rm{\overline{\nu}_{S1}=30\pm5}$) than the same value in S2s ($\rm{\overline{\nu}_{S1}=8\pm8}$, see Table \ref{tab:meanANN}).
\item The L2, T2, and SB-AGN objects have similar X-ray spectra (see Table \ref{tab:meanANN}), despite the fact that they show different spectral signatures at optical wavelengths.
\end{itemize}

In summary, our results show that the ANN is able to distinguish six classes of objects, based in their X-ray spectral shape: S1, S1.8, S2, L1.8, L2/T2/SB-AGN, and SB. One of the main differences among them is the contribution of the SB-like component, which increases as follows: $\rm{S1\Rightarrow S1.8\Rightarrow S2/L1.8 \Rightarrow}$ L2/T2/SB-AGN $\rm{\Rightarrow SB}$. 

Furthermore, apart from the Seyfert classes, the L2/T2/SB-AGN X-ray class of objects show a non-zero S1.8 component ($\rm{\overline{\nu}_{S1.8}\simeq 16}$) in their X-ray spectra, while the L1.8 class shows non-zero S1 component ($\rm{\overline{\nu}_{S1}\simeq 30}$). Therefore, our results are consistent with the hypothesis that, on average, all emission line nuclei in nearby galaxies host an AGN component, albeit of small strength in many of them. 

\begin{figure}
\centering 
\includegraphics[width=1.\columnwidth]{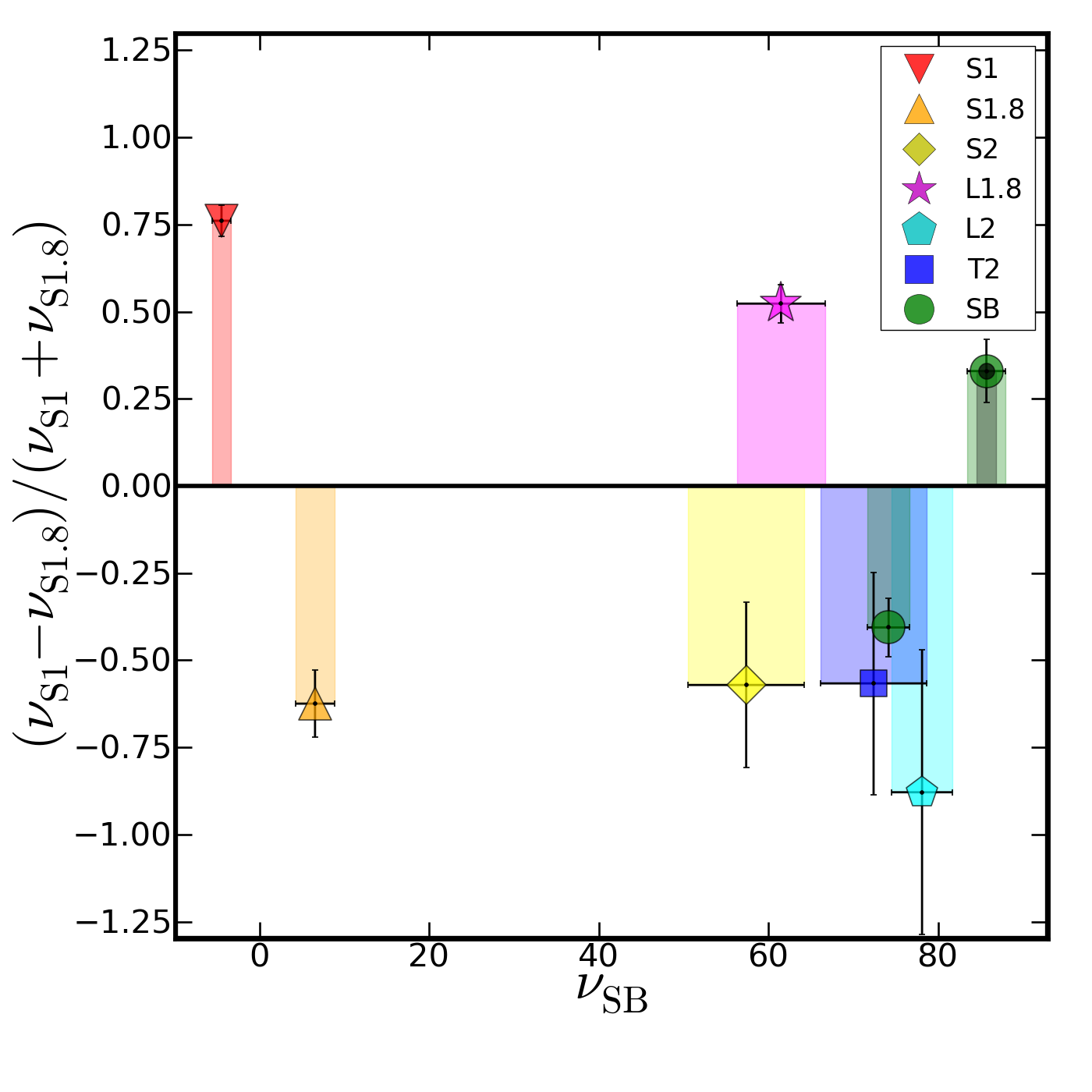}
\caption{Ratio of the difference between $\rm{\nu_{S1}}$ and $\rm{\nu_{S1.8}}$ over ($\rm{\nu_{S1}}$+ $\rm{\nu_{S1.8}}$) \emph{versus} $\rm{\nu_{SB}}$. The optical classes are shown as: S1 (red up-side down triangles), S1.8 (orange triangles), S2 (yellow diamonds), L1.8 (purple stars), L2 (light blue pentagons), T2 (dark blue squares), SB-AGN (green circles), and SB (green circles with small black dots). }
\label{fig:ANNhist2}
\end{figure}

In order to better discriminate among the classes, we built the diagram seen in Fig.~\ref{fig:ANNhist2} that shows $\rm{(\nu_{S1}-\nu_{S1.8})/(\nu_{S1}+\nu_{S1.8})}$ versus $\rm{\nu_{SB}}$. Positive (negative) values of $\rm{(\nu_{S1}-\nu_{S1.8})/(\nu_{S1}+\nu_{S1.8})}$ are expected for classes similar to the S1 (S1.8) class. The L1.8 class is similar to the S1 class with larger $\rm{\overline{\nu}_{SB}}$ than the S1 class. The S2, L2, T2, and SB-AGN classes are like the S1.8 class but with larger $\rm{\overline{\nu}_{SB}}$ than this class. The S2 class shows $\rm{\overline{\nu}_{SB}}$ similar to that of the L1.8 class. The L2, T2, and SB-AGN classes are indistinguishable. The SB class shows positive values of $\rm{(\nu_{S1}-\nu_{S1.8})/(\nu_{S1}+\nu_{S1.8})}$ with the largest $\rm{\overline{\nu}_{SB}}$ among the optical classes.

\begin{figure*}
\centering 
\includegraphics[width=1.8\columnwidth]{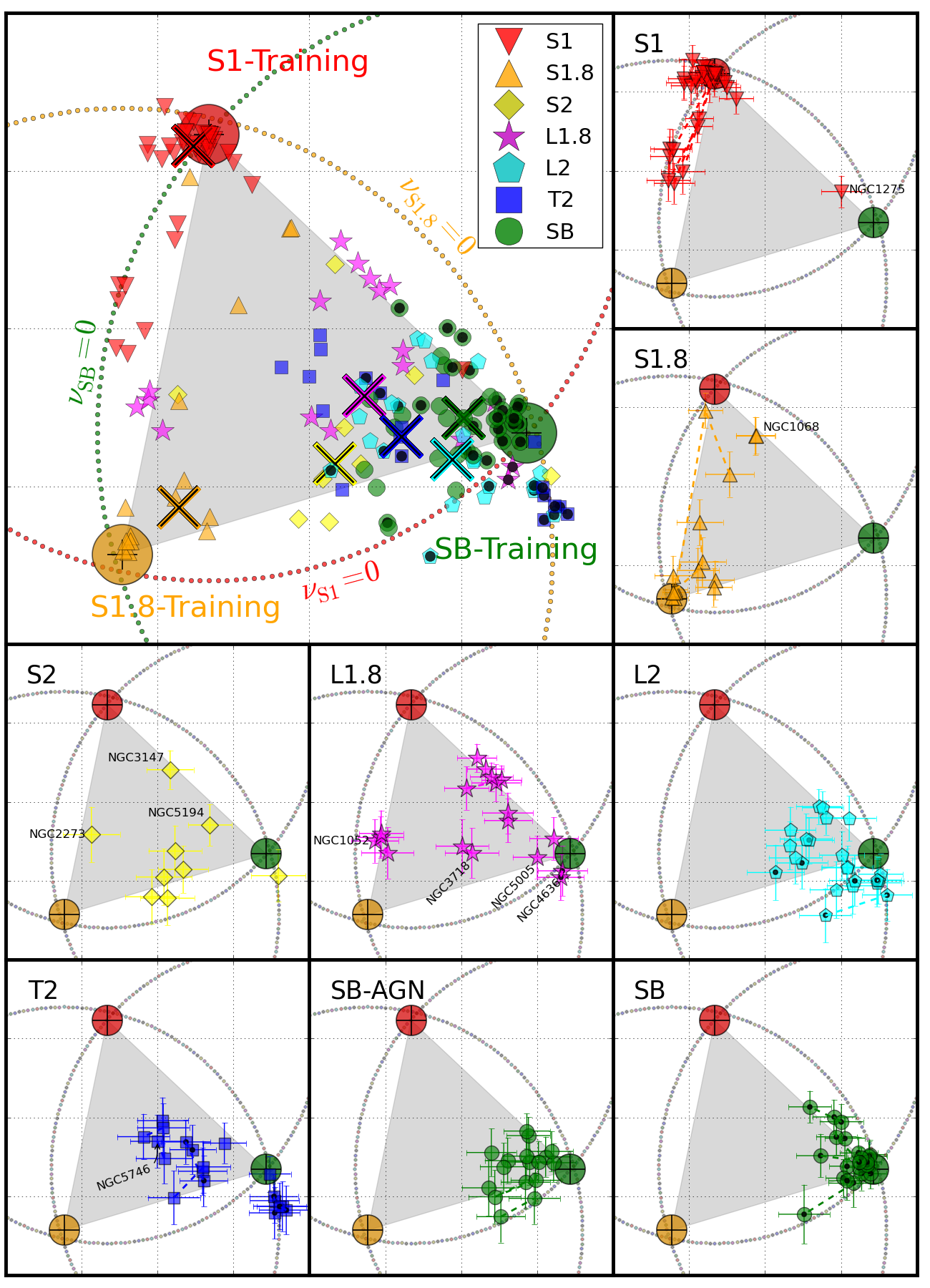}
\caption{Diagram of the ANN results. The corners of the triangle show the locus expected for S1-Training (red large circle),  S1.8-Training (orange large circle), and SB-Training (green large circle). The red, orange, and green dotted circles (centred at the corners of the triangles) correspond to $\rm{\nu_{S1}=0}$, $\rm{\nu_{ S1.8}=0}$, and $\rm{\nu_{SB}=0}$. The optical classes are shown as: S1 (red up-side down triangles), S1.8 (orange triangles), S2 (yellow diamonds), L1.8 (purple stars), L2 (light blue pentagons), T2 (dark blue squares), SB-AGN (green circles), and SB (green circles with small black dots). Black dots indicate those objects that might not be AGN according to the literature (see Table \ref{tab:sample}). Large crosses represent the mean locus for each optical class. The smaller plots show the diagrams for each optical class. Dashed lines connect observations of the same source. We have marked the names of the objects relevant for Sections \ref{sec:ANNplane} and \ref{sec:ANNcorr}. } 
\label{fig:ANNdiag}
\end{figure*}

\subsection{The ANN components plane}\label{sec:ANNplane}

All the objects in our sample are described by a combination of the three vectors of the ANN, whose sum in most cases is close to 100 including the error bars (i.e., $\rm{\nu_{S1}+\nu_{S1.8}+\nu_{SB}\simeq 100}$). Taking advantage of this, Fig. \ref{fig:ANNdiag} shows the diagram of the ANN components, plotted on a plane with these axes. The corners of the triangle represent the locus for the S1-Training, S1.8-Training, and SB-Training classes. The lines connecting each pair of these points indicate the locus on this plane for which the third component is zero. 

The ANN plane is not uniformly filled. Instead, object tend to occupy specific areas of this plane, which are distinctive of each class. This is another way to show that the the X-ray spectral shape for objects in a particular class is similar in all of them, and at the same time, these X-ray spectra are different among the various optical classes.

Most objects in the S1 and S1.8 classes are spread along a line that connects the S1- and S1.8-Training locus (except NGC\,1068 and NGC\,1275). The SB-class of objects occupy the lower right part of the diagram, close to the SB-Training locus. The objects in the L2 and T2 classes also occupy the same part of the diagram. Thus, the X-ray spectra of L2, T2, and SB-AGN are similar and close to the pure SB class (already mentioned in the previous section). Objects in the L1.8 class are spread along the line which connects the S1- and the SB-Training locus. Objects belonging to the S2 class are spread along the line connecting the S1.8- and SB-Training locus. 

\subsection{Correlations for the ANN components}\label{sec:ANNcorr}

Motivated by the results reported in the previous section regarding the position of the objects in each class in the ANN plane, we investigated the correlations between pairs of the ANN parameters. In this way, we basically project the ANN plane onto the ``$\rm{\nu_{S1.8}}$-$\rm{\nu_{S1}}$'', ``$\rm{\nu_{SB}}$-$\rm{\nu_{S1.8}}$'', and ``$\rm{\nu_{SB}}$-$\rm{\nu_{S1}}$'' relations. 

Fig. \ref{fig:ANNnuvsnu} shows $\rm{\nu_{S1.8}}$ versus $\rm{\nu_{S1}}$ (top row), $\rm{\nu_{SB}}$ versus $\rm{\nu_{S1.8}}$ (middle row), and $\rm{\nu_{SB}}$ versus $\rm{\nu_{S1}}$ (bottom row). The dashed lines in each plot indicate the locus of points for which the sum of the two ANN components is equal to 100. If an object lies on this line, then its spectrum can be reproduced by a combination of only the two ANN components relevant for each plot. For example, the X-ray spectra of the objects which are located on the diagonal line of the $\rm{\nu_{S1.8}}$ versus $\rm{\nu_{S1}}$ plot should be reproduced by a combination of {\it only} the S1.8 and S1 average X-ray spectra. Likewise for the objects located close to the dashed lines of the other panels. 

For each object in our sample, we computed the distance of each pair of its ANN components from the respective diagonal line, and we placed this object in the panel where this distance gets its smallest value. The first major result from Fig. \ref{fig:ANNnuvsnu} is that most objects in our sample are located very close (i.e., within the errors) to a diagonal line in one of the panels of this figure. This implies that the X-ray spectra in our sample are fully consistent with a combination of only {\it two} ANN components. 

Most of the objects belonging to the S1 and S1.8 classes are located in the top-left panel in Fig. \ref{fig:ANNnuvsnu}. Moreover, rather than being located around the $\rm{\nu_{S1}}$ or $\rm{\nu_{S1.8}}$, they show a continuous range of values along this diagonal line. Among the other optical classes, only NGC\,1052 and NGC\,2273 are placed in the same locus. Thus, these two sources, despite the optical classification, behave at X-rays as the S1 and S1.8 classes in our sample. 

There are no other objects, from any class, located in the $\rm{\nu_{S1.8}}$-$\rm{\nu_{S1}}$ diagonal line, except NGC\,5746. Although this source is classified as T2 by \citet{Ho97}, our results imply that its X-ray spectrum is very similar to that of S1 and S1.8 objects. Apart from this exception, the X-ray spectra of all emission line nuclei other than S1 and S1.8 classes show the contribution of a component which does not appear in the S1 and S1.8 classes. 

Most of the objects belonging to the S2 class are located along the line that connects $\rm{\nu_{SB}}$ and $\rm{\nu_{S1.8}}$ with, on average $\rm{\nu_{SB}<60}$ and little contribution from $\rm{\nu_{S1}}$ (except NGC\,2273, NGC\,5194 and NGC\,3147, Fig. \ref{fig:ANNnuvsnu}, middle row, left panel). Thus, they are similar to the S1.8 class but with larger contributions of the $\rm{\nu_{SB}}$ component. 

Most of the objects belonging to the L1.8 class fall in the $\rm{\nu_{SB}}$ versus $\rm{\nu_{S1}}$ line (except NGC\,1052, NGC\,3718, NGC\,4636, and NGC\,5005). L2 and T2 classes are placed in the same locus in these diagrams. Thus, according to the ANN, L2 and T2 classes belong to the same category. Most of them are closer to the line that connects $\rm{\nu_{SB}}$ and $\rm{\nu_{S1.8}}$ (Fig. \ref{fig:ANNnuvsnu}, middle row, middle panel) although some of them are located along the line that connects $\rm{\nu_{SB}}$ and $\rm{\nu_{S1}}$  (Fig. \ref{fig:ANNnuvsnu}, bottom row, middle panel). Moreover, a few spectra of these T2 objects are those located slightly at larger distance from the diagonal line, although still consistent with it. SB-AGN seem to be located also along the line connecting $\rm{\nu_{SB}}$ and $\rm{\nu_{S1.8}}$ (Fig. \ref{fig:ANNnuvsnu}, middle row,  right panel). Finally, most of the SB objects are located in the diagonal line connecting $\rm{\nu_{SB}}$ and $\rm{\nu_{S1}}$ (Fig. \ref{fig:ANNnuvsnu}, bottom row,  right panel). 

Based on the fact that most of the X-ray spectra in our sample can be regarded as a combination of two ANN components, we present the following scheme for the classification, based on their average X-ray spectra:

\begin{figure} 
\centering 
\includegraphics[width=1.\columnwidth]{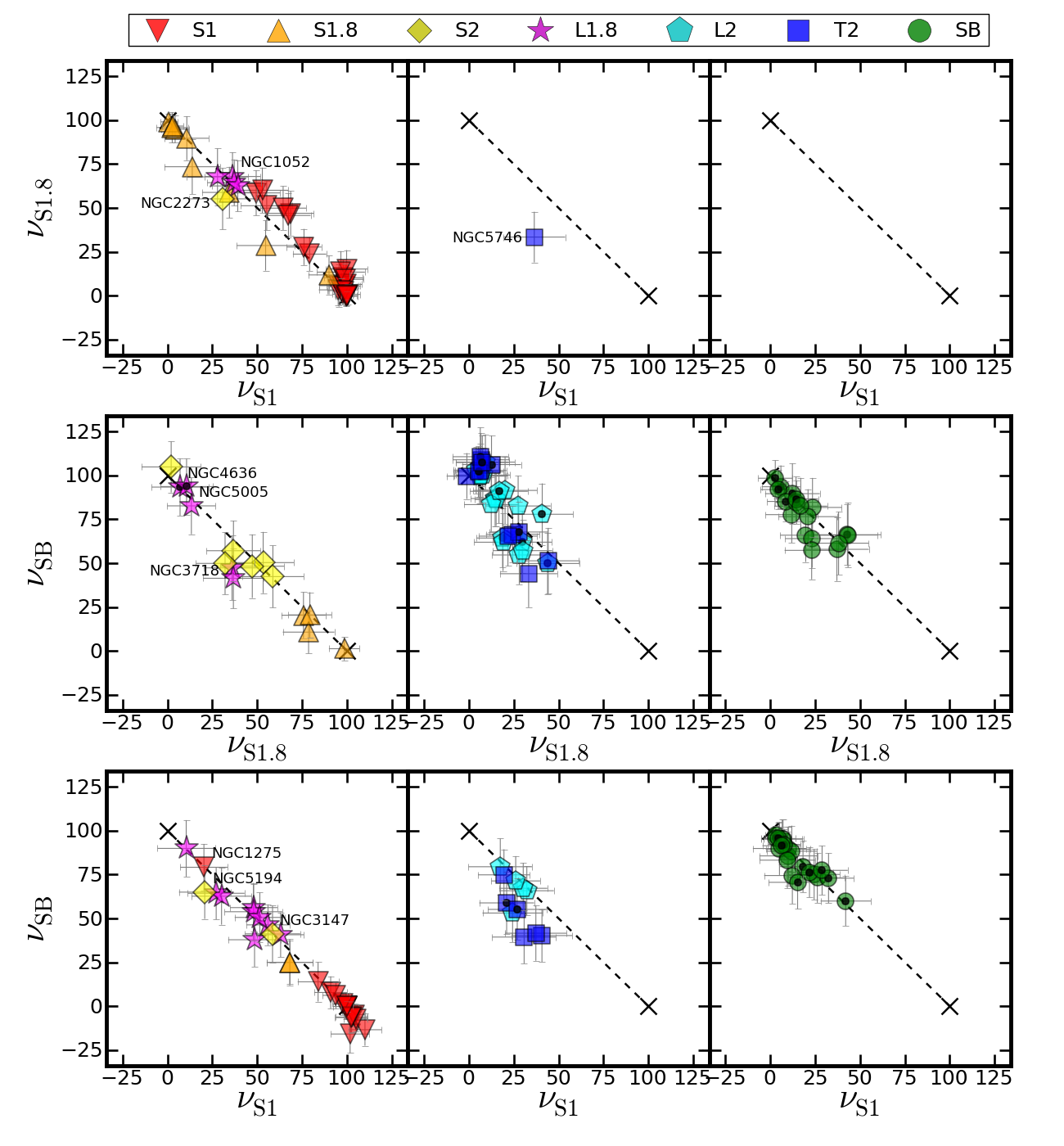}
\caption{ANN components $\nu_{S1.8}$ versus $\nu_{S1}$ (top row), $\nu_{SB}$ versus $\nu_{S1.8}$ (middle row), and $\nu_{SB}$ versus $\nu_{S1}$ (bottom row). Each row splits into three panels for the S1, S1.8, S2, and L1.8 classes (left), L2 and T2 classes (middle) and SB class (right). Dashed line shows the expected locus if the component not involved in the plot is negligible. Each plot shows those objects that are closer to its dashed line than to the dashed line of the other two plots. The optical classes are shown as: S1 (red up-side down triangles), S1.8 (orange triangles), S2 (yellow diamonds), L1.8 (purple stars), L2 (light blue pentagons), T2 (dark blue squares), SB-AGN (green circles), and SB (green circles with small black dots). Black dots indicate those objects that might not be AGN according to the literature (see Table \ref{tab:sample}).}
\label{fig:ANNnuvsnu}
\end{figure}

\begin{itemize}
\item {\bf S1 and S1.8}: They show no $\rm{\nu_{SB}}$ component ($\rm{\overline{\nu}_{SB}=-1.7\pm1.2}$ for S1 and S1.8 classes together). Large values of the $\rm{\nu_{S1}}$ component are found for the S1 class and large values of the $\rm{\nu_{S1.8}}$ component for the S1.8 class (see Table \ref{tab:meanANN}). The X-ray spectra of the objects in these two classes do show a mixture of the $\rm{\nu_{S1}}$ and $\rm{\nu_{S1.8}}$ components with a wide range of values (see Fig. \ref{fig:ANNnuvsnu}, top-left panel). 
\item {\bf S2}: They show negligible $\rm{\nu_{S1}}$ within the one sigma deviation. Their X-ray spectra are a combination of the $\rm{\nu_{SB}}$ and the $\rm{\nu_{S1.8}}$ components. 
\item {\bf L1.8}: The contribution of the $\rm{\nu_{SB}}$ component resembles that of the S2 class (see Table \ref{tab:meanANN}). However, they show larger contribution of the $\rm{\nu_{S1}}$ component compared to the S2 class. 
\item {\bf L2/T2/SB-AGN}: This family of objects shows almost no $\rm{\nu_{S1}}$ component ($\rm{\overline{\nu}_{S1}=4.1\pm2.5}$), a strong $\rm{\nu_{SB}}$ component ($\rm{\overline{\nu}_{SB}=75.1\pm2.6}$) and a smaller $\rm{\nu_{S1.8}}$ component ($\rm{\overline{\nu}_{S1.8}=17.0\pm1.6}$). It can be distinguished from the S2 class because of their significantly larger mean value of the $\rm{\nu_{SB}}$ component (see Table \ref{tab:meanANN}). 
\item {\bf SB}: This is the class of objects that show the largest values for the $\rm{\nu_{SB}}$ component ($\rm{\overline{\nu}_{SB}=86 \pm 2.}$) and almost non-existent $\rm{\nu_{S1}}$ ($\rm{\overline{\nu}_{S1}=8.7 \pm 2.7}$) and $\rm{\nu_{S1.8}}$ components ($\rm{\overline{\nu}_{S1.8}= 4.4 \pm 2.3}$).   
\end{itemize}

\section{Discussion}\label{sec:discussion}

We have shown that the ANN analysis can be useful to classify the main optical classes using only X-ray spectra. In general, an object with $\rm{\nu_{SB}\le10}$ is almost certainly a S1 or a S1.8. Moreover, an object with little $\rm{\nu_{S1.8}}$ and large $\rm{\nu_{S1}}$ and $\rm{\nu_{SB}}$ is most probably a L1.8, while an object with little $\rm{\nu_{S1}}$ and large $\rm{\nu_{S1.8}}$ and $\rm{\nu_{SB}}$ is most probably a S2. Larger fractions of $\rm{\nu_{SB}}$ characterise the L2,T2 and SB nuclei. However, we would like to stress that most of the differences are found when we consider the average value for each class. Thus, although we believe that the ANN method is very useful to study the average properties, it may not be as successful in classifying a single object based on its ANN components. Using the results regarding the average properties of the objects in each class, in this section we discuss the following questions: (1) Type-1/Type-2 dichotomy; (2) optical versus X-ray classes; and (3) elusive AGN. Finally, we present the utility of this analysis for its application to X-ray surveys. 

\begin{figure} 
\centering 
\includegraphics[width=1.\columnwidth]{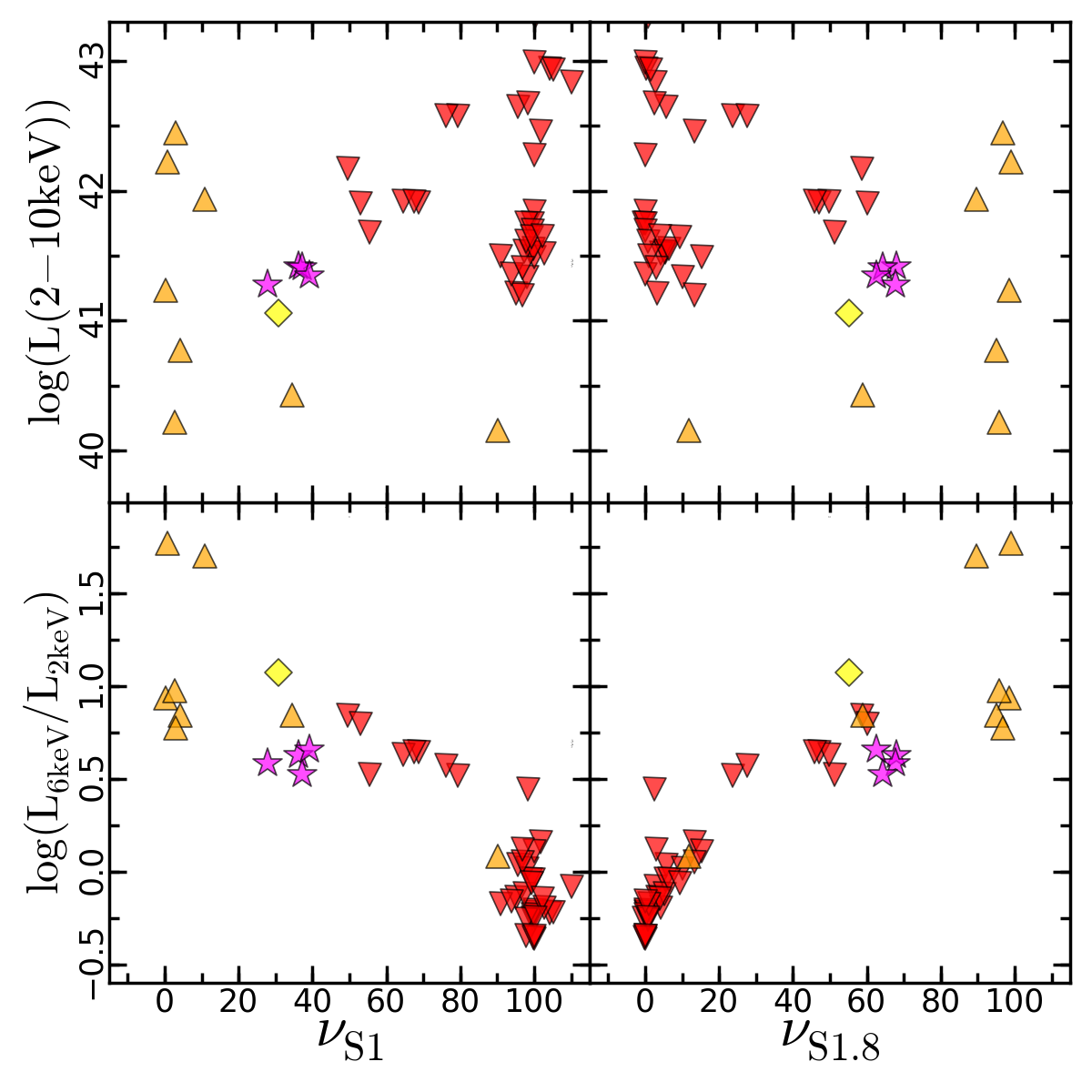}
\caption{The logarithmic of the 2-10 keV band observed luminosity, log(L(2-10 keV)), versus the $\rm{\nu_{S1}}$ (left) and $\rm{\nu_{S1.8}}$ (right) components. (Bottom): The logarithmic of the ratio between the observed luminosity at 6 keV versus the observed luminosity at 2 keV, $\rm{log(L_{6keV}/L_{2keV})}$, versus the $\rm{\nu_{S1}}$ (left) and $\rm{\nu_{S1.8}}$ (right) components. We only plot objects with $\rm{\nu_{SB}<10}$ (see text).  The optical classes are shown as: S1 (red up-side down triangles), S1.8 (orange triangles), S2 (yellow diamonds), and L1.8 (purple stars).}
\label{fig:MeanANN3vsLsteep}
\end{figure}

\subsection{Type-1/Type-2 dichotomy}\label{sec:dichotomy}

Our results indicate that the X-ray spectra of the S1 and S1.8 classes can be reproduced by a mixture of the $\rm{\nu_{S1}}$ and $\rm{\nu_{S1.8}}$ components, with $\rm{\nu_{S1}}$ and $\rm{\nu_{S1.8}}$ being stronger in the former and latter classes, respectively. Furthermore, the S1 and S1.8 classes show a continuous range of values of the $\rm{\nu_{S1}}$ and $\rm{\nu_{S1.8}}$ components (see Fig. \ref{fig:ANNnuvsnu}, top-left panel). Our analysis cannot offer direct indications of the nature of the $\rm{\nu_{S1}}$ or the $\rm{\nu_{S1.8}}$ components, or for the physical parameter that drives their correlation for S1s and S1.8s. Below we discuss possible interpretations of this result.

The continuous range of values for $\rm{\nu_{S1}}$ and $\rm{\nu_{S1.8}}$ could reflect a continuous range of absorptions (i.e., $\rm{N_H}$), increasing for the S1.8 class. This is consistent with the UM of AGN. Indeed, X-rays have been used in AGN to study the amount of absorption \citep{Risaliti99,Bianchi12,Ho08}. \citet{Risaliti99} found that 75\% of their Type-2 Seyferts were heavily obscured ($\rm{N_{H}>10^{23}cm^{-2}}$), 50\% of them were Compton-thick (i.e., $\rm{N_{H}>1.5\times 10^{24}cm^{-2}}$), with the S1.8 class characterised by an average lower $\rm{N_H}$ than the S2 class. Alternatively, a low flux level continuum is recently suggested by \citet{Elitzur14} as the main reason to classify objects as S1.8s. They suggest that intermediate types of objects are part of an evolutionary sequence where the BLR slowly disappears as the bolometric luminosity decreases. Hence, the continuous range of values for $\rm{\nu_{S1}}$ and $\rm{\nu_{S1.8}}$ could be interpreted either as (1) an increase of the absorption as we move from S1s and S1.8s or (2) a decrease of the AGN continuum flux in S1.8s. As shown below, our results favour the first interpretation. 

Assuming that L(2-10 keV) is an indication of the total luminosity, we would expect it to be proportional to $\rm{\nu_{S1}}$ and inversely correlated with $\rm{\nu_{S1.8}}$ if a decrease of the intrinsic continuum is responsible for the S1.8 class. Fig. \ref{fig:MeanANN3vsLsteep} (top panels) shows the log(L(2-10 keV))\footnote{L(2-10 keV) is computed as the sum of all the bins in the calibrated spectra in the 2-10 keV band multiplied by the size of the spectral bin ($\rm{\Delta E = 0.05}$ keV).} versus $\rm{\nu_{S1}}$ (left) and $\rm{\nu_{S1.8}}$ (right) for objects with a negligible contribution of $\rm{\nu_{SB}}$ ($\rm{\nu_{SB}<10}$).  At each $\rm{\nu_{S1}}$ or $\rm{\nu_{S1.8}}$ values there is a large scatter of luminosities, but objects with large (small) $\rm{\nu_{S1}}$ ($\rm{\nu_{S1.8}}$) have larger X-ray luminosities, on average. The Pearson's correlation coefficients are r=0.37 ($\rm{P_{null}=0.008}$) and r=0.34 ($\rm{P_{null}=0.015}$) for the correlations with $\rm{\nu_{S1}}$ and $\rm{\nu_{S1.8}}$, respectively (see Fig. \ref{fig:MeanANN3vsLsteep}, top, right and left panels). The small numbers of the correlation coefficients shows that the correlations are not strong, although the null hypothesis probability indicates that it may be significant.  

The bottom panels of Fig. \ref{fig:MeanANN3vsLsteep} show the steepness of the spectra, expressed as $\rm{log(L_{6keV}/L_{2keV})}$\footnote{$\rm{L_{2keV}}$ and $\rm{L_{6keV}}$ are the monochromatic luminosities at 2 keV and 6 keV, respectively, obtained from the flux-calibrated spectra.}, versus the $\rm{\nu_{S1}}$ (left) and $\rm{\nu_{S1.8}}$ (right) components. The X-ray spectra become harder (i.e., the emission at 6 keV becomes more prominent compared to the emission at 2 keV) when the $\rm{\nu_{S1.8}}$ component increases (and $\rm{\nu_{S1}}$ decreases). The correlation between them shows Pearson's correlation coefficients and null probabilities of r=0.91, $\rm{P_{null}=6\times10^{-20}}$ and r=0.89, $\rm{P_{null}=7\times10^{-18}}$, respectively. 

Irrespective of the reason for the spectral hardening, the strength of the correlations in the lower panels of Fig. \ref{fig:MeanANN3vsLsteep} indicates that the distributions of the $\rm{\nu_{S1}}$ and $\rm{\nu_{S1.8}}$ components in Seyferts are not driven, primarily, by luminosity, but by the spectral hardening of their X-ray spectra. The simplest explanation for this spectral hardening is an increase of absorption, which in the case of Compton-thin sources affects much stronger the 2 keV flux than the 6 keV flux. Therefore, based on the strength of the correlations shown in Fig. \ref{fig:MeanANN3vsLsteep} it seems reasonable to assume that a variable amount of obscuration is the main physical parameter responsible for the continuous range of $\rm{\nu_{S1}}$ and $\rm{\nu_{S1.8}}$. The same effect can also explain the weak correlations with the luminosity (see Fig. \ref{fig:MeanANN3vsLsteep}, top panels). If the observed luminosities are corrected for absorption, then both S1.8 and S1 could show the same level of X-ray luminosity. Therefore, we believe that the scenario to be preferred is that in which obscuration is responsible for the Type-1/Type-2 dichotomy. This is fully consistent with the UM of AGN, in which the obscuring torus is responsible for blocking the inner parts of the AGN (both the BLR and the X-ray source) in Type-2 galaxies.

\begin{figure} 
\centering 
\includegraphics[width=1.\columnwidth]{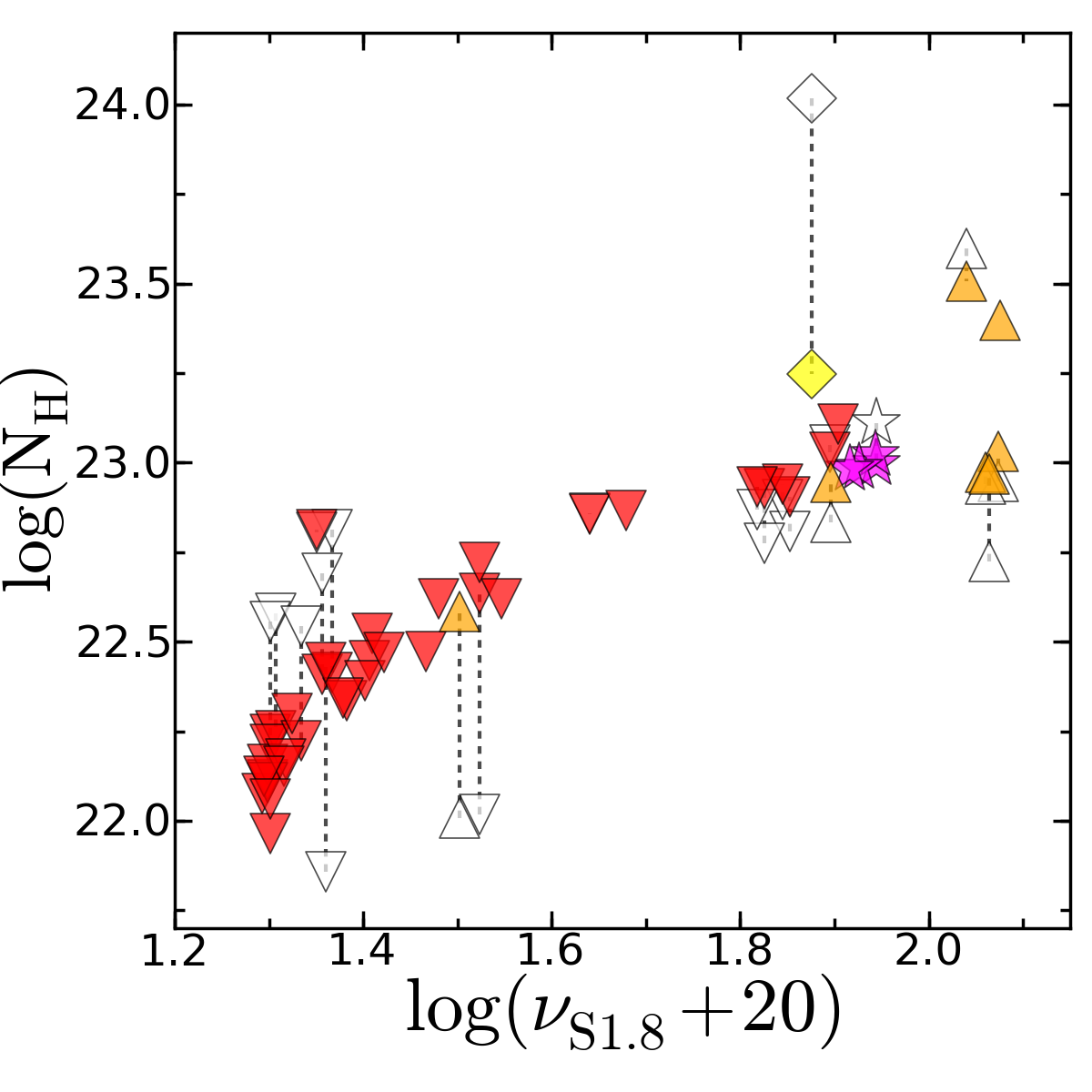}
\caption{The logarithmic of the $\rm{N_H}$ versus $\rm{log(\nu_{S1.8}+20)}$. Filled symbols show $\rm{N_H}$ values using a simple power-law model to the 2-10 keV band (see Appendix A). Empty symbols show $\rm{N_H}$ reported in the literature when available (see Table \ref{tab:NHs}). Dashed vertical lines link the $\rm{N_H}$ values using a simple power-law model and those reported in the literature.}
\label{fig:NHvsANN}
\end{figure}

A final check on the nature of this dichotomy can be performed comparing $\rm{\nu_{S1.8}}$ with the absorbing column density, $\rm{N_H}$, for these observations (see Fig. \ref{fig:NHvsANN} and Appendix A for the details on the measurements of $\rm{N_H}$). The quantity $\rm{log(\nu_{S1.8}+20)}$\footnote{Note that we have computed the logarithmic of $\rm{(\nu_{S1.8}+20)}$ in order to avoid negative values of $\rm{\nu_{S1.8}}$.} is linearly related with $\rm{log(N_H)}$ (r=0.93, $\rm{P_{null}=1.5\times10^{-21}}$) when derived with a simple power-law fit (filled symbols in Fig. \ref{fig:NHvsANN}). A less significant linear relation (r=0.57, $\rm{P_{null}=5.3\times10^{-3}}$) is found when using $\rm{N_H}$ estimates reported in the literature (empty symbols in Fig. \ref{fig:NHvsANN}). We believe this weaker relationship is due to: (1) less number of observations with $\rm{N_H}$ and (2) different models used for the spectral fittings for each observation. It reinforces the importance on the do a self-consistent modelling for the sample to compare the parameters.

\subsection{Optical versus X-ray classes}\label{sec:BPT}

The ANN has found differences on the average X-ray spectra of the six different classes: S1, S1.8, S2, L1.8, L2/T2/SB-AGN, and SB. Thus, the L2, T2, and SB-AGN belong to the same X-ray category according to the ANN results. It is worth to remark that division lines in the BPT diagrams have been developed and adapted as a function of the ionisation models and/or observations available \citep[e.g.,][]{Veilleux87,Osterbrock89,Kewley01,Kauffmann03,Kewley06,Stasinska06,Kewley13}. Objects close to the division between star-forming galaxies and AGN could be classified as L2, T2, or SB depending on how these divisions are set and/or how these three diagrams are used together. This could explain why the L2, T2, and SB-AGN classes cannot be distinguished at X-rays according to the ANN. Alternatively, the number of physical parameters governing the classes at X-rays might be lower than those driving the optical classes. 

\begin{figure} 
\centering 
\includegraphics[width=1.\columnwidth]{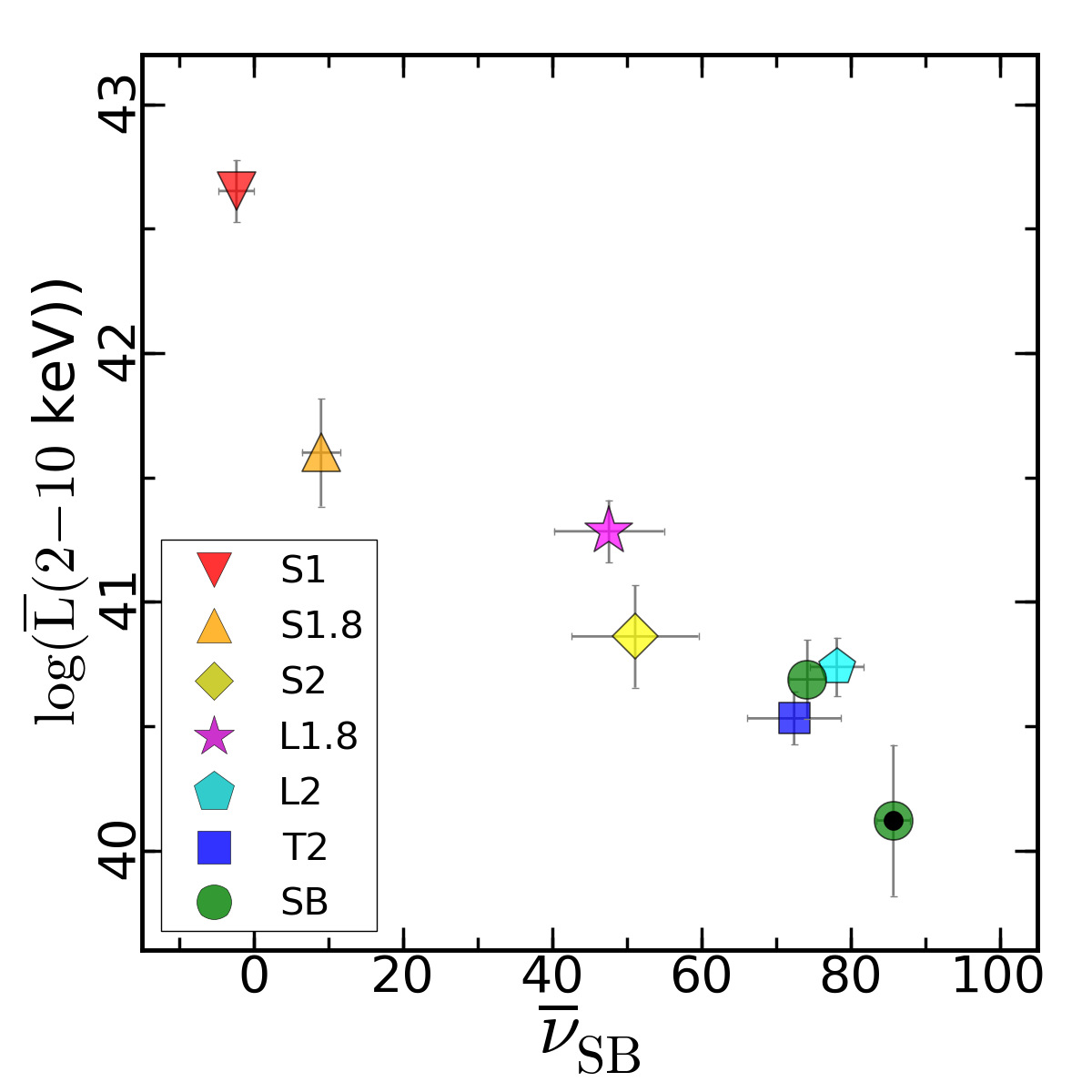}
\caption{The mean $\rm{\overline{\nu}_{SB}}$ component versus the mean 2-10 keV band observed luminosity in logarithmic scale, $\rm{log(\overline{L}(2-10~keV))}$, per optical class. The optical classes are shown as: S1 (red up-side down triangles), S1.8 (orange triangles), S2 (yellow diamonds), L1.8 (purple stars), L2 (light blue pentagons), T2 (dark blue squares), SB-AGN (green circles), and SB (green circles with small black dots).}
\label{fig:MeanANN3vsLhard}
\end{figure}

One of the main differences between the X-ray spectra of the various optical classes is set by the $\rm{\nu_{SB}}$ component, which is increasing from the S1 to the SB classes, passing through the S1.8, S2, L1.8, and L2/T2/SB-AGN groups. The nature of the $\rm{\nu_{SB}}$ component cannot be fully assessed with the results of this analysis alone, but we discuss below possible explanations. 

The star-formation (circumnuclear or that of the host galaxy) is the most natural explanation for the $\rm{\nu_{SB}}$ component. In this case, X-ray emission by binary systems, supernovae remnants, and/or emission by diffuse hot gas, could contribute to this $\rm{\nu_{SB}}$ component. In this case we would expect $\rm{\overline{\nu}_{SB}}$ to increase when the luminosity decreases for the objects in our sample. To test such hypothesis Fig. \ref{fig:MeanANN3vsLhard} shows the average $\rm{\nu_{SB}}$ ($\rm{\overline{\nu}_{SB}}$) versus the mean value for log(L(2-10 keV)). These two quantities are in fact clearly anti-correlated (r=0.94, $\rm{P_{null}=5\times10^{-5}}$)\footnote{Note that $\rm{\nu_{S1}}$ and $\rm{\nu_{S1.8}}$ show a poor relation with $\rm{log(\overline{L}(2-10 keV))}$ in the top panel of Fig. \ref{fig:MeanANN3vsLsteep}.}. Thus, $\rm{\nu_{SB}}$ increases when the X-ray luminosity decreases, in favour of our hypothesis that the $\rm{\nu_{SB}}$ component is related to star-formation. The SB galaxies, with the largest $\rm{\nu_{SB}}$ values in our sample, have an X-ray luminosities of $\rm{\sim 10^{40} erg~s^{-1}}$. This could be representative of the galactic X-ray emission due to the processes mentioned above. If an AGN component is present in almost all galaxies, then as it becomes stronger, $\rm{\nu_{SB}}$ decreases, while at the same time the X-ray luminosity increases. The $\rm{\nu_{SB}}$ component is almost zero in the S1.8 and S1 classes probably because the AGN-like source entirely outshines the underlying host-galaxy emission, or it could also mean that the $\rm{\nu_{SB}}$ component is entirely absent. For example, \citet{Wu09} (and references therein) claimed that the circumnuclear star-formation might be even destroyed in the presence of an AGN.  

An alternative origin for the $\rm{\nu_{SB}}$ component for those sources hosting an AGN is the X-ray emission from the hot plasma in the NLR, emission from the ``scattering component'' in AGN or ionised gas. It has been claimed that high resolution X-ray spectra are dominated by emission lines from the NLR in Type-2 Seyferts \citep{Guainazzi07}. Moreover, the soft X-ray emission in a few AGN is extended on scales ranging from a few hundred parsecs to a few thousand parsecs, in close agreement with the morphology of the NLR seen at optical wavelengths for both LINERs and Type-2 Seyferts \citep[][]{Gonzalez-Martin10,Bianchi06,Masegosa11}. In this case we would expect $\rm{\overline{\nu}_{SB}}$ to increase when the luminosity increases for the objects in our sample. However, as mentioned before, $\rm{\nu_{SB}}$ increases when the X-ray luminosity decreases (see Fig. \ref{fig:MeanANN3vsLhard}), ruling out the NLR as the main responsible for the $\rm{\nu_{SB}}$ component.

\subsection{Elusive AGN}\label{sec:hidden}

The $\rm{\nu_{S1}}$ and/or $\rm{\nu_{S1.8}}$ components are not negligible in most of the emission line nuclei presented in this paper (see Figs. \ref{fig:ANNhist} and \ref{fig:ANNnuvsnu}). A total of 22 out of the 162 spectra (i.e., 13.5\%) are consistent with no signature of an AGN-like component; this percentage is slightly higher in terms of the number of objects (19 out of the 90, 21\%). Thus, $\rm{\sim}$80\% of our sample show signs of an AGN-like component, either with a S1-like or a S1.8-like contribution. This number is almost twice the percentage of AGN (43\%) estimated at optical frequencies by \citet{Ho97} for the same sample. Moreover, although for some T2, SB-AGN, and SB nuclei, the $\rm{\nu_{S1}}$ and $\rm{\nu_{S1.8}}$ components are consistent with zero, on average, the X-ray spectra of these nuclei do show the presence of $\rm{\nu_{S1}}$ or $\rm{\nu_{S1.8}}$ components. However, based on their optical spectra, these classes correspond, at best, to objects on the border between AGN and star-forming galaxies. 

Our result strongly supports the hypothesis that an AGN component might be present at X-rays at a certain level in most of the emission line nuclei included in our sample, even if they do not show signatures of this AGN component in their optical spectra. Non-AGN at optical wavelengths with AGN signatures at X-rays have been largely studied in the literature \citep[called `elusive AGN', see][]{Maiolino98,Soria06A,Soria06B}. Galaxies with bulges harbour BHs \citep[see][and references therein]{Kormendy13}. However, at optical wavelengths, only a small fraction of bulge galaxies show evidence for AGN activity; in about half of the high signal-to-noise (S/N) ratio optical spectra taken by \citet{Ho97} there is no indication of AGN activity. \citet{Tzanavaris07} studied a sample of star-forming galaxies classified by \citet{Ho97} at X-ray, finding AGN signatures for a large fraction of them. This is consistent with our results. \citet{Tzanavaris07} suggested that the lack of optical signatures may be due to the fact that the emission could be overwhelmed by that coming from circumnuclear star formation. This is entirely consistent with the increase of the $\rm{\nu_{SB}}$ component when the luminosity decreases (see Fig. \ref{fig:MeanANN3vsLhard} and previous Section), if the $\rm{\nu_{SB}}$ component is associated with the constant, diffuse X-ray emission of the host galaxy and/or X-ray emission associated with intense star-forming regions. 

\subsection{Relevance of the ANN method for X-ray surveys}\label{sec:surveys}

ANNs have proven to be a powerful approach to a broad variety of problems \citep[e.g.,][]{Bishop96, Gupta04, Asensio-Ramos05, Socas-Navarro05, Carballo08, Han12}. In the most common application, ANN functions as a classification algorithm. In the AGN field, for instance, \citet{Rawson96} already used the ANN to classify optical spectra into Type-1 and Type-2 AGN. However, ANN have not been used to classify X-ray spectra before. 

Using other statistical methods, several attempts have been made to classify X-ray spectra, particularly for low S/N spectra. \citet{Norman04} selected normal, Type-1 and Type-2 AGN galaxies from the \emph{Chandra} Deep field North (CDF-N) and South (CDF-S) samples using a Bayesian classification procedure. Priors were constructed from a set of galaxies with well-defined optical classes. They used the X-ray hardness ratio, the 0.5-2 keV X-ray luminosity, and the ratio between X-ray and optical fluxes. The product of the prior distribution for a class and the likelihood for the observed parameters for a given source gave the probability that the source was drawn from that class. \citet{Ptak07} used a similar methodology with several improvements (e.g., \emph{k}-correction in the optical data). They showed that the method was efficient in classifying the X-ray spectra into Type-1, Type-2 and normal galaxies. Our methodology has two advantages: (1) it does not need any optical information and (2) it is able to distinguish among the S1, S1.8, L1.8, S2, L2/T2/SB-AGN, and SB classes. We show that the ANN is an excellent tool to discriminate between most of the optical classes using only their X-ray spectra. It might be very useful for X-ray surveys where the optical information is missed. The ANN components can be computed for any set of X-ray spectra using our already trained ANN\footnote{We kindly suggest to contact any of the coauthors of the paper for the use of our trained ANN.}. The effects of using X-ray spectra with lower S/N to their classification with the ANN method needs to be explored (perhaps through simulations), which is out of the scope of this paper. Finally, the ANN should be able to classify objects in broad classes, and the results will be useful for statistical studies. However, the method is not being demonstrated to be particularly useful in the classification of objects on an individual basis.

\section{Summary}\label{sec:summary}

We have investigated the connection between optical classes and X-ray spectra in a sample of 90 nearby emission line galaxies. We have used flux-calibrated X-ray spectra observed with \emph{XMM}-Newton/pn. The results of this paper are, for the first time, free of the subjectivity of the X-ray spectral fitting thanks to the use of the ANNs:

\begin{itemize}

\item We used a set of the S1, S1.8 and SB classes to train the ANN, giving as output arrays $\rm{\nu_{S1}}$, $\rm{\nu_{S1.8}}$, and $\rm{\nu_{SB}}$, respectively. The ANN is 90\% efficient to distinguish these classes. They all show then distinctive signatures at X-rays. 

\item Based on their X-ray spectral shape, the emission line nuclei in the nearby galaxies are divided into six groups: S1, S1.8, S2, L1.8, L2/T2/SB-AGN, and SB classes. Only the L2, T2, and SB-AGN classes show the same average X-ray spectrum even though they belong to distinct optical classes. Furthermore, the objects within each of these six classes have similar average X-ray spectra. 

\item  The average X-ray spectrum of the objects in each X-ray class can be described by the contribution of two components, either the $\rm{\nu_{SB}}$ and $\rm{\nu_{S1}}$, the $\rm{\nu_{SB}}$ and $\rm{\nu_{S1.8}}$, or the $\rm{\nu_{S1}}$ and $\rm{\nu_{S1.8}}$ (in the case of S1s and S1.8s). The S2 (L1.8) class is similar to the S1.8 (S1) class but with larger contributions of the $\rm{\nu_{SB}}$ component. The L2/T2 and SB-AGN classes have a strong $\rm{\nu_{SB}}$ component, with the addition of a $\rm{\nu_{S1.8}}$ component.

\item The S1 and S1.8 classes show little $\rm{\nu_{SB}}$ and a wide range of the $\rm{\nu_{S1}}$ and $\rm{\nu_{S1.8}}$ components. We show that this wide range of $\rm{\nu_{S1}}$ and $\rm{\nu_{S1.8}}$ contributions is most probably related to the different amount of obscuration that affects the nuclear emission at X-rays, in agreement with the UM predictions.

\item Most of the objects in our sample have a non negligible contribution of either a $\rm{\nu_{S1}}$ or a $\rm{\nu_{S1.8}}$ component. This result strongly supports the presence of an AGN-like nucleus in most nearby galaxies, albeit at different levels of luminosities (i.e. activity).  

\item We argue that the $\rm{\nu_{SB}}$ component is associated to a contribution of star-formation in the host-galaxy. As the contribution of the AGN component decreases, the $\rm{\nu_{SB}}$ component increases, and at optical wavelengths it shows stronger signatures representative of S2, L1.8, L2/T2/SB-AGN, and finally of SB nuclei. 

\end{itemize}

We find that the emission line nuclei in nearby galaxies can be classified in six classes, based on the shape of their X-ray spectra. These classes are associated to the traditional optical classes, although their number is smaller. Thus, the shape of the X-ray spectra of those galaxies may be determined by a smaller number of physical parameters than those which determine the optical classes. Alternatively, this could be due to the difficulties to classify them at optical wavelength using the BPT diagrams. Indeed, our results suggest that the X-ray spectra of nearby galaxies are simply the combination of two components. The first one is an AGN-like component and the second one is due to star-formation in the host-galaxy contributing to the X-rays. An AGN-like nucleus may be present in most of them (80\%). Its strength, relative to the contribution of star-formation in the host-galaxy, determines the average X-ray spectrum of objects for each X-ray class. A third physical parameter could be related to the amount of obscuring material along the LOS. This parameter almost certainly drives the Type-1/Type-2 dichotomy, but may also explain why, for example, the L1.8 class predominantly shows a $\rm{\nu_{S1}}$ component in their spectra while L2, T2, and SB-AGN predominantly show a $\rm{\nu_{S1.8}}$ component.

We conclude that the ANN method is quite powerful to detect AGN-like nuclei (and distinguish which ones are affected by absorption). It can therefore be used to identify AGN, and even to infer their optical classes, using only X-ray spectra and our trained ANN. However, this can only be done in a statistical way, i.e., using the X-ray spectra of a large number of objects. This methodology could be very useful in X-ray surveys, e.g. the \emph{eRosita} survey, where the optical information for tens of thousands newly discovered objects will not be available.

\begin{acknowledgements} 
We thank to the anonymous referee for his/her useful comments and suggestions. The authors acknowledge the Spanish MINECO through project Consolider-Ingenio 2010 Program grant CSD2006-00070: First Science with the GTC (http://www.iac.es/consolider-ingenio-gtc/) and AYA2012-39168-C03-01. This work was also partially funded by the Spanish MINECO through a Juan de la Cierva Fellowship.This work was financed by MINECO grant AYA 2010-15169, Junta de Andaluc\'ia TIC114 and Proyecto de Excelencia de la Junta de Andaluc\'ia P08-TIC-03531. LHG also thanks to the grant BES-2011-043319 for her financial support. Based on observations obtained with \emph{XMM}-Newton, an ESA science mission with instruments and contributions directly funded by ESA Member States and NASA. This research has made use of data obtained from the High Energy Astrophysics Science Archive Research Center (HEASARC), provided by NASA's Goddard Space Flight Center.
\end{acknowledgements}

\scriptsize{
\onecolumn
\setcounter{table}{0}
\begin{longtable}{r l c c c c c c r r r r r r r r}
\hline\hline  	       
    &  	      &	       	  &           &         &             &    & \multicolumn{3}{c}{ANN} & Comments  \\	\cline{7-10} 
    &  	      &	       	  &	&     &         &       	  &		&	       &	      & 	   &		      \\
Num &   Name  &    ObsID  &   Class  &  Expos. &Counts   &  Train     & $\rm{\nu_{S1}}$  &  $\rm{\nu_{S1.8}}$ & $\rm{\nu_{SB}}$ &  \\
(1) & (2)     &   (3)     &    (4)   &   (5)   &  (6)  &   (7)                  &   (8)             & (9)      	  &      (10)    &   (11)	 \\ \hline			  
    &  	      &	       	  &	     &         &       &       			&                   & 		  &		 &  	     		   \\
\endfirsthead
\hline\hline                 
    &  	      &	       	  &           &         &             &    & \multicolumn{3}{c}{ANN} & Comments  \\	\cline{7-10}
    &  	      &	       	  &	&     &         &       	  &		&	       &	      & 	   &		      \\
Num &   Name  &    ObsID  &   Class  &  Expos. &Counts   &  Train     & $\rm{\nu_{S1}}$  &  $\rm{\nu_{S1.8}}$ & $\rm{\nu_{SB}}$ &  \\
(1) & (2)     &   (3)     &    (4)   &   (5)   &  (6)  &   (7)                  &   (8)             & (9)      	  &      (10)    &   (11)	 \\ \hline			  
    &  	      &	       	  &	     &         &       &       			&                   & 		  &		 &  	     		   \\
\endhead
    1&           IC10&  152260101&   H&   30&	15238&   -- &	18.2$\pm$  13.3 &   -2.7$\pm$  13.1 &	79.5$\pm$  14.7 &   \\
    2&          IC342&  093640901&   H&    4&	 1283&   TR &	-0.6$\pm$  11.2 &    5.9$\pm$  10.7 &	93.6$\pm$  12.2 &   \\
     &               &  206890201&    &   16&	 3108&   -- &	-5.3$\pm$  16.6 &   12.3$\pm$  15.3 &	89.6$\pm$  17.3 &   \\
     &               &  206890401&    &    3&	  744&   -- &  -14.2$\pm$  16.9 &   43.0$\pm$  18.7 &	66.2$\pm$  17.4 &   \\
    3&         NGC315&  305290201&L1.9&   13&	 5785&   -- &	10.5$\pm$  16.0 &    0.0$\pm$  13.5 &	89.9$\pm$  16.1 &  AGN (1) \\
    4&         NGC410&  203610201& T2:&   13&	 5848&   -- &  -25.3$\pm$  16.7 &    6.7$\pm$  15.8 &  108.9$\pm$  15.2  &  Non-AGN (1) \\
     &               &  304160201&    &    5&	 2400&   -- &  -26.6$\pm$  16.7 &   12.6$\pm$  16.4 &  106.2$\pm$  16.5 &   \\
    5&         NGC598&  141980501&   H&    1&	 9053&   -- &	 9.9$\pm$  12.4 &   -1.3$\pm$  10.3 &	89.5$\pm$  12.6 &   \\
     &               &  141980801&    &    7&	34989&   TR &	 3.0$\pm$   5.6 &   -0.3$\pm$	4.2 &	97.4$\pm$   5.5 &   \\
     &               &  102640101&    &    5&	33527&   -- &	 7.1$\pm$  10.8 &   -4.9$\pm$	9.6 &	95.2$\pm$  11.1 &   \\
     &               &  102642101&    &    8&	25042&   -- &	 6.6$\pm$  11.8 &   -3.9$\pm$	9.4 &	95.3$\pm$  11.1 &   \\
    6&         NGC777&  304160301&S2/L&    0&	  639&   -- &  -27.8$\pm$  17.3 &   40.9$\pm$  17.2 &	77.9$\pm$  17.6 & Non-AGN $\clubsuit$  \\
     &               &  203610301&    &    3&	 3138&   -- &  -27.4$\pm$  16.5 &    9.2$\pm$  16.4 &  107.6$\pm$  15.8 &   \\
    7&        NGC1052&  553300401&L1.9&   46&	30645&   -- &	36.3$\pm$  15.6 &   67.9$\pm$  13.8 &	-4.3$\pm$   8.3 & AGN (1)  \\
     &               &  093630101&    &   11&	 5954&   -- &	27.9$\pm$  17.1 &   67.7$\pm$  16.2 &	 3.4$\pm$  12.4 &   \\
     &               &  553300301&    &   42&	27269&   -- &	37.2$\pm$  15.1 &   64.4$\pm$  13.2 &	-1.7$\pm$   8.2 &   \\
     &               &  306230101&    &   44&	25545&   -- &	39.2$\pm$  14.5 &   62.5$\pm$  14.6 &	-1.8$\pm$   9.4 &   \\
    8&        NGC1068&  111200101&S1.8&   32&  368177&   -- &	68.3$\pm$  12.7 &    2.7$\pm$  12.1 &	24.9$\pm$  12.3 & CT (2)  \\
     &               &  111200201&    &   27&  316879&   -- &	68.2$\pm$  12.9 &    2.2$\pm$  11.7 &	25.0$\pm$  13.2 &   \\
    9&        NGC1275&  085110101&S1.5&   22&  431300&   -- &	20.3$\pm$  13.2 &   -0.6$\pm$  10.0 &	79.2$\pm$  13.3 &   \\
   10&        NGC1569&  112290801&   H&   11&	 2178&   TR &	-2.1$\pm$  10.2 &    2.8$\pm$	9.5 &	98.4$\pm$  10.2 &   \\
   11&        NGC2146& 110930101&   H&    6&	 3430&   -- &	10.0$\pm$  16.1 &    4.0$\pm$  12.7 &	85.7$\pm$  16.8 &  AGN (3) \\
   12&        NGC2273&  140951001&  S2&    3&	  455&   -- &	30.8$\pm$  18.7 &   55.2$\pm$  17.5 &	 0.9$\pm$  18.1 &  CT (2,4) \\
   13&        NGC2342&  093190501&   H&   23&	 2532&   TR &	 3.0$\pm$  11.4 &    4.5$\pm$  10.0 &	92.0$\pm$  11.3 &   \\
   14&        NGC2655&  301650301&  S2&    1&	  590&   -- &  -10.2$\pm$  17.7 &   53.4$\pm$  17.1 &	50.3$\pm$  17.5 &  AGN (1)  \\
   15&        NGC2787&  200250101&L1.9&   25&	 2511&   -- &	27.2$\pm$  15.8 &   10.3$\pm$  13.9 &	64.3$\pm$  15.0 &  AGN (1) \\
   16&        NGC2841&  201440101&  L2&    9&	 1731&   -- &	16.8$\pm$  15.5 &   28.1$\pm$  15.2 &	54.7$\pm$  16.7 & AGN (1)   \\
   17&        NGC2903&  556280301&   H&   54&	15114&   TR &	 3.1$\pm$   6.9 &    0.6$\pm$	5.8 &	95.7$\pm$   7.1 &  \\
   18&        NGC3079&  110930201&  S2&    4&	 1132&   -- &	 3.9$\pm$  16.3 &   36.8$\pm$  15.0 &	57.1$\pm$  17.1 & CT (2,4)  \\
   19&        NGC3147&  405020601&  S2&   12&	 8073&   -- &	58.6$\pm$  15.4 &    2.9$\pm$  12.2 &	41.2$\pm$  16.2 & True-S2 (5)	\\
   20&        NGC3226&  101040301&L1.9&   30&	 6518&   -- &	48.6$\pm$  14.7 &   12.3$\pm$  13.7 &	37.8$\pm$  15.4 & AGN (1)  \\
     &               &  400270101&    &   93&	28221&   -- &	48.1$\pm$  13.2 &   -3.7$\pm$	9.5 &	56.2$\pm$  13.4 &   \\
   21&        NGC3227&  101040301&S1.5&   30&	32042&   -- &	55.4$\pm$  14.4 &   51.2$\pm$  14.0 &	-6.0$\pm$   8.2 &   \\
     &               &  400270101&    &   93& 1184930&   TR &  100.0$\pm$   2.5 &    0.0$\pm$	2.1 &	-0.0$\pm$   2.5 &   \\
   22&        NGC3310&  556280101&   H&   21&	15781&   -- &	24.4$\pm$  14.3 &    2.4$\pm$  11.7 &	75.1$\pm$  14.4 &  AGN (3) \\
     &               &  556280201&    &   24&	16355&   -- &	26.2$\pm$  15.2 &    0.8$\pm$  11.2 &	73.1$\pm$  14.2 &   \\
   23&        NGC3367& 551450101&   H&    9&	 1572&   -- &	 9.7$\pm$  16.0 &   11.9$\pm$  14.6 &	77.7$\pm$  16.0 & AGN (6)   \\
   24&        NGC3516&  107460601&S1.2&   43&  249895&   -- &	95.6$\pm$  11.0 &    5.7$\pm$  12.0 &	-1.9$\pm$  10.1 &   \\
     &               &  107460701&    &   81&  299978&   -- &  101.8$\pm$  10.7 &   13.4$\pm$  12.5 &  -15.9$\pm$  10.8 &   \\
     &               &  401210501&    &   40& 1112450&   -- &  104.2$\pm$   5.9 &    0.3$\pm$	5.5 &	-4.9$\pm$   6.3 &   \\
     &               &  401210601&    &   42&  589580&   -- &  110.1$\pm$   9.2 &    2.7$\pm$	9.4 &  -13.4$\pm$   9.1 &   \\
     &               &  401210401&    &   30&  880682&   TR &  100.0$\pm$   2.4 &    0.0$\pm$	2.1 &	-0.0$\pm$   2.5 &   \\
     &               &  401211001&    &   35&  933142&   -- &  105.1$\pm$   6.6 &    1.6$\pm$	5.9 &	-7.3$\pm$   7.0 &   \\
   25&        NGC3623&  082140301& L2:&   25&	 2873&   -- &	16.7$\pm$  16.5 &   18.8$\pm$  14.6 &	64.3$\pm$  16.7 & Non-AGN (1)  \\
   26&        NGC3628&  110980101&  T2&   37&	 6061&   -- &	40.7$\pm$  16.9 &   19.3$\pm$  14.7 &	40.2$\pm$  15.1 & AGN (7)  \\
   27&        NGC3665& 052140201&  H:&   20&	 2777&   -- &	13.8$\pm$  15.1 &   19.7$\pm$  14.3 &	66.0$\pm$  15.3 &  AGN (8) \\
   28&       NGC3690A& 679381101&   H&    6&	 4058&   -- &	 5.1$\pm$  14.8 &    3.0$\pm$  12.1 &	89.9$\pm$  15.1 & CT (2,4)  \\
     &               &  112810101&    &   13&	 8136&   -- &	12.0$\pm$  14.1 &   -0.4$\pm$  12.8 &	88.3$\pm$  14.6 &   \\
   29&        NGC3718&  200430501&L1.9&    9&	 2479&   -- &	14.5$\pm$  17.6 &   36.0$\pm$  15.8 &	47.3$\pm$  18.3 & AGN (1)  \\
     &               &  200431301&    &    7&	 1696&   -- &	19.3$\pm$  17.3 &   36.8$\pm$  16.9 &	41.4$\pm$  17.2 &   \\
   30&        NGC3884&  301900601&L1.9&   18&	 7065&   -- &	47.9$\pm$  16.1 &   -0.4$\pm$  11.7 &	53.8$\pm$  16.4 & AGN (1) \\
   31&        NGC3998&  090020101&L1.9&    5&	35635&   -- &	56.1$\pm$  12.1 &   -2.2$\pm$	8.7 &	46.3$\pm$  11.7 & AGN (1)  \\
   32&        NGC4051&  157560101&S1.2&   40&  233043&   -- &	95.2$\pm$  10.8 &    3.3$\pm$	8.9 &	 0.3$\pm$  10.4 &   \\
     &               &  606321601&    &   24&  850091&   TR &  100.0$\pm$   2.2 &    0.0$\pm$	2.0 &	 0.0$\pm$   2.2 &   \\
     &               &  606320201&    &   16&  360147&   -- &	99.4$\pm$   7.1 &   -0.4$\pm$	5.9 &	-0.2$\pm$   7.4 &   \\
     &               &  606321901&    &   13&	80791&   -- &	98.0$\pm$  10.8 &   10.2$\pm$  10.4 &	-8.5$\pm$  11.0 &   \\
     &               &  606322001&    &    8&	79259&   -- &	98.9$\pm$   8.7 &    5.5$\pm$	8.7 &	-4.6$\pm$   9.6 &   \\
     &               &  606320301&    &   13&  335449&   -- &	97.8$\pm$   5.8 &   -0.1$\pm$	5.7 &	 1.3$\pm$   6.4 &   \\
     &               &  606320401&    &   12&	78135&   -- &	99.9$\pm$  11.6 &   15.2$\pm$  10.8 &  -15.4$\pm$  11.7 &   \\
     &               &  606321501&    &   10&  227170&   -- &	99.5$\pm$   6.0 &   -0.1$\pm$	5.7 &	-1.2$\pm$   6.8 &   \\
     &               &  606321701&    &   21&  183454&   -- &  102.7$\pm$   8.8 &    4.1$\pm$	7.1 &	-6.2$\pm$   8.8 &   \\
     &               &  606322301&    &   22&  317693&   -- &	97.2$\pm$   8.4 &    5.2$\pm$	8.5 &	-3.1$\pm$   8.1 &   \\
     &               &  606322201&    &   19&  192988&   -- &	99.9$\pm$   8.6 &    6.4$\pm$	8.6 &	-6.7$\pm$   9.7 &   \\
     &               &  606321401&    &   24&  454259&   -- &	97.9$\pm$   5.8 &    0.8$\pm$	5.7 &	-0.0$\pm$   6.4 &   \\
     &               &  606322101&    &   16&	62470&   -- &	96.7$\pm$  13.4 &   13.4$\pm$  12.0 &  -11.9$\pm$  13.5 &   \\
     &               &  606321801&    &   11&  129550&   -- &	99.2$\pm$  10.2 &    9.3$\pm$	8.8 &  -10.3$\pm$  11.5 &   \\
     &               &  606320101&    &   27&  319242&   -- &  102.2$\pm$   8.7 &    3.9$\pm$	6.7 &	-6.5$\pm$   8.8 &   \\
   33&        NGC4102& 601780701&   H&    5&	 1293&   -- &	 1.7$\pm$  17.7 &   37.1$\pm$  17.8 &	57.9$\pm$  18.1 &  AGN (9) \\
   34&        NGC4138&  112551201&S1.9&    8&	 4253&   TR &	 0.2$\pm$   8.9 &   98.5$\pm$	8.5 &	 1.2$\pm$   6.7 &    \\
   35&        NGC4143&  150010601&L1.9&    9&	 3153&   -- &	30.2$\pm$  16.8 &    6.9$\pm$  12.9 &	62.7$\pm$  16.2 & AGN (1)  \\
   36&        NGC4151&  112310101&S1.5&   20&  138931&   -- &	68.7$\pm$  12.9 &   46.9$\pm$  12.7 &  -15.4$\pm$   8.9 &   \\
     &               &  112830501&    &   17&  115289&   -- &	64.5$\pm$  12.9 &   49.9$\pm$  12.5 &  -15.2$\pm$   8.8 &   \\
     &               &  112830201&    &   50&  330369&   -- &	67.5$\pm$  12.3 &   45.7$\pm$  12.7 &  -14.3$\pm$   9.1 &   \\
     &               &  143500301&    &   12&  381987&   TR &	98.2$\pm$   4.7 &    2.4$\pm$	4.8 &	-0.0$\pm$   2.2 &   \\
     &               &  402660201&    &   21&  197054&   -- &	49.5$\pm$  13.4 &   58.6$\pm$  12.9 &  -10.4$\pm$   7.7 &   \\
     &               &  143500201&    &   12&  294178&   -- &	79.2$\pm$   9.3 &   23.7$\pm$	9.6 &	-3.7$\pm$   5.0 &   \\
     &               &  143500101&    &    9&  232041&   -- &	76.1$\pm$   9.7 &   27.7$\pm$  10.3 &	-3.1$\pm$   5.3 &   \\
     &               &  402660101&    &   25&  157409&   -- &	52.9$\pm$  14.5 &   60.1$\pm$  12.9 &  -12.8$\pm$  10.5 & \\
   37&        NGC4157&  203170101&   H&   30&	 2583&   TR &	 5.7$\pm$  11.3 &    8.7$\pm$  10.4 &	85.2$\pm$  11.3 &  \\
   38&        NGC4168&  112550501&S1.9&   15&	 1053&   TR &	 2.4$\pm$  12.5 &   75.8$\pm$  12.3 &	20.3$\pm$  13.2 & True-S2 (5) \\
   39&        NGC4214&  035940201&   H&    9&	 1290&   TR &	-0.1$\pm$  11.7 &   14.6$\pm$  11.9 &	84.7$\pm$  12.4 &   \\
   40&        NGC4235&  204650201&S1.2&    8&	 8137&   TR &	91.0$\pm$   9.1 &    1.1$\pm$	7.6 &	 7.8$\pm$   9.2 &   \\
   41&        NGC4254& 147610101&   H&   11&	 1901&   -- &	12.3$\pm$  15.7 &   10.5$\pm$  14.8 &	74.5$\pm$  16.1 &  AGN (10) \\
   42&        NGC4258&  400560301&S1.9&   43&	35016&   -- &	34.6$\pm$  15.2 &   58.8$\pm$  14.6 &	 6.4$\pm$  11.1 &   \\
     &               &  059140901&    &    9&	 7884&   -- &	13.8$\pm$  15.5 &   73.5$\pm$  15.6 &	11.9$\pm$  12.4 &   \\
     &               &  059140101&    &    7&	 7752&   -- &	10.5$\pm$  14.1 &   78.7$\pm$  14.4 &	10.7$\pm$  12.1 &   \\
     &               &  110920101&    &   11&	14231&   TR &	 4.3$\pm$   7.9 &   94.9$\pm$	7.6 &	 1.1$\pm$   5.5 &   \\
   43&        NGC4261&  056340101&  L2&   21&	10740&   -- &	-3.7$\pm$  16.8 &   14.6$\pm$  13.6 &	86.9$\pm$  15.8 &  AGN (1) \\
     &               &  502120101&    &   63&	32201&   -- &	-4.0$\pm$  16.6 &   15.2$\pm$  12.0 &	86.3$\pm$  15.1 &   \\
   44&        NGC4278&  205010101&L1.9&   20&	34532&   -- &	51.5$\pm$  14.0 &   -1.1$\pm$	9.2 &	50.6$\pm$  14.2 & AGN(1)  \\
   45&        NGC4303& 205360101&   H&   15&	 2624&   -- &	10.6$\pm$  15.7 &   23.3$\pm$  15.0 &	64.0$\pm$  16.6 &  AGN (6) \\
   46&        NGC4314&  201690301&  L2&   14&	 1275&   -- &	 5.0$\pm$  16.8 &   29.8$\pm$  16.4 &	62.0$\pm$  17.7 & Non-AGN (1)  \\
   47&        NGC4321&  106860201&  T2&    9&	 2663&   -- &	 8.7$\pm$  16.1 &   24.4$\pm$  15.0 &	66.5$\pm$  16.1 & AGN (10)  \\
   48&        NGC4374&  673310101&  L2&   24&	19790&   -- &  -14.4$\pm$  14.7 &    4.1$\pm$  12.8 &  103.0$\pm$  13.4 & CT (11)  \\
   49&        NGC4378&  301650801&  S2&    9&	  605&   -- &	-6.7$\pm$  17.7 &   58.8$\pm$  17.2 &	42.4$\pm$  17.7 &  True-S2 (5)  \\
   50&        NGC4388&  110930701&S1.9&    6&	 9801&   -- &	10.8$\pm$  12.2 &   89.7$\pm$  12.6 &	-2.5$\pm$  10.6 &   \\
     &               &  675140101&    &   21&	57292&   TR &	 0.9$\pm$   4.7 &   99.1$\pm$	4.5 &	-0.0$\pm$   3.3 &   \\
   51&        NGC4395&  112522701&S1.8&    5&	 6908&   -- &	54.7$\pm$  16.1 &   28.6$\pm$  14.3 &	17.2$\pm$  13.2 &   \\
     &               &  142830101&    &   86&  100135&   -- &	90.1$\pm$  11.5 &   11.8$\pm$  11.1 &	-3.2$\pm$  10.7 &   \\
     &               &  112521901&    &    8&	 5453&   TR &	 2.7$\pm$   8.8 &   95.8$\pm$	8.7 &	 1.2$\pm$   6.3 &   \\
   52&        NGC4414&  402830101& T2:&   16&	 3047&   -- &	21.1$\pm$  15.3 &   19.5$\pm$  15.6 &	58.9$\pm$  17.5 & Non-AGN $\clubsuit$  \\
   53&        NGC4459&  550540101& T2:&   61&	 6839&   -- &	27.0$\pm$  16.4 &   19.4$\pm$  14.0 &	55.2$\pm$  17.5 & Non-AGN (1)	\\
     &               &  550540201&    &   15&	 1543&   -- &	 3.2$\pm$  15.6 &   27.7$\pm$  16.2 &	67.8$\pm$  17.1 &   \\
   54&        NGC4472&  200130101&S2::&   72&  115802&   -- &  -17.3$\pm$  17.9 &    2.1$\pm$  16.5 &  104.9$\pm$  14.6 &  True-S2 (5) \\
   55&        NGC4486&  114120101&  L2&   24&  361682&   -- &	17.5$\pm$  17.7 &   -7.0$\pm$  14.2 &	79.5$\pm$  16.1 &  AGN (1)  \\
   56&        NGC4490&  112280201&   H&   11&	 2399&   -- &	15.2$\pm$  16.1 &   14.4$\pm$  13.8 &	70.6$\pm$  15.2 &   \\
     &               &  556300101&    &   18&	 5260&   TR &	 4.5$\pm$   9.5 &   -0.7$\pm$	7.5 &	95.6$\pm$   8.8 &   \\
   57&        NGC4494&  071340301&L2::&   23&	 2101&   -- &	24.4$\pm$  16.8 &   21.6$\pm$  14.8 &	53.1$\pm$  16.6 &  AGN (1) \\
   58&        NGC4526& 205010201&   H&   18&	 2739&   -- &	21.4$\pm$  16.4 &   23.0$\pm$  15.1 &	57.3$\pm$  16.6 &  AGN (12) \\
   59&        NGC4552&  141570101& T2:&   17&	12193&   -- &	-6.0$\pm$  13.8 &   -1.1$\pm$  11.3 &	99.7$\pm$  13.1 & AGN (1,10 )  \\
   60&        NGC4559&  152170501&   H&   33&	 9950&   TR &	 7.7$\pm$   8.9 &    0.0$\pm$	6.4 &	92.2$\pm$   8.9 &    \\
   61&        NGC4565&  112550301&S1.9&    8&	 1236&   TR &	-0.9$\pm$  12.0 &   79.4$\pm$  12.0 &	20.5$\pm$  12.9 &  True-S2 (5) \\
   62&        NGC4569&  200650101&  T2&   41&	 6720&   -- &	19.5$\pm$  14.5 &    5.0$\pm$  12.9 &	74.9$\pm$  14.6 &  AGN (10) \\
   63&        NGC4579&  112840101&S1.9&   14&	37493&   -- &	63.1$\pm$  13.0 &   -3.2$\pm$	9.1 &	40.7$\pm$  12.7 &  AGN (1,10) \\
   64&        NGC4594&  084030101&  L2&   15&	11642&   -- &	32.3$\pm$  15.0 &    1.6$\pm$  11.1 &	65.8$\pm$  14.1 & AGN (1,10)  \\
   65&        NGC4636&  111190701&L1.9&   50&	91742&   -- &  -12.7$\pm$  16.4 &    7.1$\pm$  15.9 &	93.4$\pm$  16.3 & Non-AGN (1)  \\
     &               &  111190201&    &    5&	 9876&   -- &  -15.1$\pm$  15.3 &   10.7$\pm$  14.6 &	93.7$\pm$  16.0 &   \\
   66&        NGC4639&  112551001&S1.0&    8&	 2826&   TR &	84.0$\pm$  11.4 &    1.9$\pm$	9.5 &	13.9$\pm$  11.3 &   \\
   67&        NGC4698&  651360401&  S2&   27&	 1722&   -- &	 3.5$\pm$  18.3 &   47.3$\pm$  17.8 &	48.3$\pm$  18.1 & CT (11)  \\
   68&        NGC4725&  112550401& S2:&   11&	 1333&   -- &	13.4$\pm$  17.4 &   32.1$\pm$  16.2 &	49.9$\pm$  17.7 & True-S2 (5)  \\
   69&        NGC4736&  404980101&  L2&   33&	42014&   -- &	30.3$\pm$  13.5 &    0.2$\pm$  10.3 &	67.2$\pm$  14.5 & AGN (1,10)  \\
     &               &  094360601&    &    8&	11071&   -- &	26.1$\pm$  13.6 &    3.7$\pm$  10.8 &	71.4$\pm$  14.2 &   \\
   70&        NGC4845&  658400601&   H&   14&	83763&   -- &	 2.9$\pm$   6.1 &   96.7$\pm$	6.0 &	-0.0$\pm$   3.2 &  AGN (13)  \\
   71&        NGC5005&  110930501&L1.9&    8&	 2952&   -- &	 2.8$\pm$  16.1 &   13.4$\pm$  13.5 &	82.7$\pm$  16.2 & CT (4,11)  \\
   72&        NGC5033&  094360501&S1.5&    8&	21897&   TR &	93.9$\pm$   7.5 &   -0.1$\pm$	5.4 &	 6.2$\pm$   7.6 &  AGN (1,10) \\
   73&        NGC5055&  405080301&  T2&    5&	 1675&   -- &	 9.5$\pm$  17.9 &   21.8$\pm$  16.3 &	65.4$\pm$  17.0 & AGN (10)  \\
     &               &  405080501&    &    2&	  812&   -- &	-3.7$\pm$  17.2 &   44.2$\pm$  17.0 &	51.3$\pm$  18.8 &   \\
   74&        NGC5194&  112840201&  S2&   17&	11622&   -- &	20.8$\pm$  14.3 &    8.4$\pm$  13.2 &	65.0$\pm$  15.3 &  CT (2,4) \\
   75&        NGC5195&  303420201& L2:&   20&	 3318&   -- &  -13.9$\pm$  17.1 &   20.2$\pm$  17.8 &	91.5$\pm$  19.3 &  AGN (10) \\
     &               &  212480801&    &   22&	 3764&   -- &	 4.3$\pm$  14.8 &   12.5$\pm$  15.1 &	83.4$\pm$  15.5 &   \\
   76&        NGC5204&  142770301&   H&    3&	 2771&   TR &	 6.2$\pm$  10.6 &    2.4$\pm$	9.1 &	91.5$\pm$  10.7 &   \\
     &               &  150650301&    &    4&	 4484&   -- &	21.8$\pm$  13.4 &    0.7$\pm$  11.6 &	76.3$\pm$  14.3 &   \\
     &               &  405690201&    &   25&	27564&   -- &	32.4$\pm$  14.1 &   -6.2$\pm$  10.9 &	73.0$\pm$  14.2 &   \\
     &               &  405690101&    &    7&	 9860&   -- &	28.8$\pm$  14.3 &   -7.1$\pm$  11.1 &	77.4$\pm$  14.0 &   \\
     &               &  405690501&    &   19&	16050&   -- &	42.0$\pm$  14.0 &   -0.9$\pm$  10.1 &	59.9$\pm$  14.2 &   \\
   77&        NGC5248&  655380401&   H&    7&	  941&   TR &	-2.2$\pm$  12.2 &   14.3$\pm$  12.3 &	86.6$\pm$  12.6 &   \\
   78&        NGC5273&  112551701&S1.5&    9&	15529&   TR &	96.8$\pm$   7.6 &    2.9$\pm$	6.9 &	 0.4$\pm$   7.1 &   \\
   79&        NGC5322&  071340501&L2::&   13&	 1344&   -- &	 7.7$\pm$  17.5 &   29.9$\pm$  16.5 &	57.2$\pm$  17.5 & AGN (14)  \\
   80&        NGC5363&  201670201&  L2&   11&	 1850&   -- &	17.1$\pm$  17.3 &   19.0$\pm$  16.1 &	62.0$\pm$  16.4 & AGN (1)  \\
   81&        NGC5548&  109960101&S1.5&   15&  309750&   -- &  100.3$\pm$   7.0 &    0.7$\pm$	5.9 &	-2.2$\pm$   7.5 &   \\
     &               &  089960301&    &   47& 1185380&   -- &	99.7$\pm$   6.2 &   -0.3$\pm$	5.1 &	 0.9$\pm$   6.1 &   \\
     &               &  089960401&    &   18&  636996&   TR &  100.0$\pm$   3.2 &    0.0$\pm$	2.4 &	 0.0$\pm$   3.1 &   \\
   82&        NGC5746&  651890101&  T2&   42&	 2986&   -- &	30.8$\pm$  17.8 &   29.5$\pm$  16.6 &	39.4$\pm$  15.1 & AGN (1)  \\
     &               &  651890201&    &   28&	 1814&   -- &	19.9$\pm$  18.7 &   33.3$\pm$  16.0 &	44.0$\pm$  19.1 &   \\
     &               &  651890301&    &   66&	 4867&   -- &	37.3$\pm$  17.2 &   22.2$\pm$  15.5 &	41.6$\pm$  15.7 &   \\
     &               &  651890401&    &   54&	 3692&   -- &	36.4$\pm$  17.2 &   33.4$\pm$  14.4 &	32.6$\pm$  15.4 &   \\
   83&        NGC5813&  302460101& L2:&   19&	25523&   -- &  -20.5$\pm$  15.8 &    7.8$\pm$  15.2 &  100.2$\pm$  15.6 & Non-AGN (1)  \\
     &               &  554680201&    &   43&	59102&   -- &  -20.9$\pm$  16.4 &    5.9$\pm$  14.7 &	99.6$\pm$  15.4 &   \\
     &               &  554680301&    &   42&	58293&   -- &  -19.4$\pm$  16.0 &    6.6$\pm$  14.9 &	99.7$\pm$  15.8 &   \\
   84&        NGC5846&  021540501& T2:&   10&	10036&   -- &  -22.9$\pm$  16.2 &    6.6$\pm$  14.3 &  102.9$\pm$  15.2 & Non-AGN (1)  \\
     &               &  021540101&    &   25&	25881&   -- &  -20.6$\pm$  15.9 &    5.3$\pm$  14.1 &  102.1$\pm$  14.5 &   \\
   85&        NGC5982&  673770401&L2::&    8&	 1920&   -- &  -12.8$\pm$  17.1 &   16.8$\pm$  16.1 &	91.0$\pm$  18.3 & Non-AGN (14)  \\
   86&        NGC6217& 400920101&   H&    6&	 1011&   -- &  -16.5$\pm$  16.0 &   43.3$\pm$  16.4 &	66.0$\pm$  18.5 & AGN (15)  \\
     &               &  400920201&    &    7&	 1137&   -- &  -12.2$\pm$  16.9 &   23.5$\pm$  16.3 &	81.8$\pm$  16.8 &   \\
   87&        NGC6482&  304160801&T2/S&    4&	 3163&   -- &  -28.6$\pm$  17.8 &    6.4$\pm$  15.5 &  110.7$\pm$  16.7 & Non-AGN (1)  \\
     &               &  304160401&    &    6&	 4811&   -- &  -25.1$\pm$  17.5 &    7.5$\pm$  14.8 &  107.4$\pm$  15.4 &   \\
   88&        NGC6703&  601830401&L2::&   14&	 1154&   -- &	 4.9$\pm$  18.9 &   43.9$\pm$  17.5 &	49.8$\pm$  17.6 &  Non-AGN $\clubsuit$ \\
   89&        NGC6946& 200670301&   H&    7&	 1108&   -- &	-1.2$\pm$  16.2 &   21.0$\pm$  16.5 &	76.5$\pm$  15.7 &   AGN (16) \\
     &               &  200670401&    &    4&	  703&   -- &	-4.6$\pm$  17.1 &   38.0$\pm$  16.8 &	61.4$\pm$  17.9 &   \\
     &               &  500730101&    &   19&	 2281&   -- &	-0.8$\pm$  16.5 &   17.0$\pm$  14.8 &	82.5$\pm$  16.1 &   \\
     &               &  500730201&    &   26&	 4205&   -- &	 9.5$\pm$  15.5 &    9.2$\pm$  13.0 &	83.2$\pm$  16.1 &   \\
   90&        NGC7626&  149240101&L2::&   33&	 3041&   -- &  -14.4$\pm$  17.7 &   27.5$\pm$  18.1 &	82.6$\pm$  17.3 & AGN (17)  \\
   \\ \hline
\caption{\scriptsize{Sample properties and results. The net number of counts in
the X-ray 0.2-10 keV band. Exposure time in ksec. References for the AGN nature
for nonsecure AGN (AGN/Non-AGN), Compton-thickness (CT), and True Type-2
Seyferts (True,S2) included in Col. 11 (see Section \ref{sec:sample}): 
(1) \citet{Gonzalez-Martin09a},
(2) \citet{Goulding12},
(3) \citet{Tzanavaris07},
(4) \citet{Comastri04},
(5) \citet{Akylas08},
(6) \citet{Veron06},
(7) \citet{Goulding09},
(9) \citet{Gonzalez-Martin11},
(10) \citet{Moustakas10},
(11) \citet{Gonzalez-Martin09b},
(12) \citet{Davis13},
(13) \citet{Nikolajuk13},
(14) \citet{Tomita00},
(15) \citet{Nicholson97},
(16) \citet{Elmegreen98}, and
(17) \citet{Randall09}. Non-AGN marked as $\clubsuit$ are those without indications of AGN activity reported in the literature.
}}
\label{tab:sample}  
\end{longtable}
\twocolumn
}

\newpage
\onecolumn

\appendix

\section{Hydrogen column densities}
\normalsize{
We have estimated the hydrogen column densities, $\rm{N_{H}}$, for the observations in our sample with $\rm{\nu_{SB}<10}$ with the purpose of comparing them with $\rm{\nu_{S1.8}}$ (see Fig. \ref{fig:NHvsANN} and Section \ref{sec:dichotomy}). The spectral fitting has been performed using XSPEC v12.7.1. The spectra were binned to a minimum of 20 counts per spectral bin before background subtraction to use the $\chi^2$ statistics. The task {\sc grppha} included in the {\sc ftools} software has been used for this purpose. We used the simplest fit that represents the data, a single power-law. Therefore, we fitted the spectra in the 2--10~keV energy band to a single power-law, with a fixed spectral index of $\rm{\Gamma=2.1}$. This power-law is attenuated by an absorber, {\sc zwabs} within Xspec, and three Gaussian profiles centred at 6.4, 6.7, and 6.95 keV are added to include the plausible existence of the neutral and ionised iron lines. Note that the width of the lines was fixed to the spectral resolution of \emph{XMM}-Newton for the ionised lines but it was let to vary for the neutral FeK$\alpha$ line. The resulting $\rm{N_{H}}$ estimates are reported in Table \ref{tab:NHs}. This table also includes the $\rm{N_{H}}$ values reported in the literature for the same observations. We only include the spectral fittings corresponding to the same observations included in our analysis because many of these objects are highly variable. Our $\rm{N_{H}}$ values agree with those from literature for large $\rm{N_{H}}$. Discrepancies up to a factor of $\rm{\sim 3}$ are found for smaller $\rm{N_{H}}$ values. The spectral fittings performed in the literature have the advantage of providing a more realistic modelling of the spectra. However, we have found very few of them and the modelling performed is not the same in all the cases. Our simple spectral fitting has the advantage of being homogenous and it is available for all the objects with $\rm{\nu_{SB}<10}$. Thus, our simple spectral fitting is better suited for the comparison of $\rm{N_{H}}$ and $\rm{\nu_{S1.8}}$ (see Fig. \ref{fig:NHvsANN}).}

\begin{table}[!h]
\centering
\scriptsize
\begin{tabular}{l c c c l}
\hline \hline
   Name     & ObsID & \multicolumn{3}{c}{$\rm{log(N_{H})}$} \\  \cline{3-5}
        &   & Own & Literature & Ref. \\ \hline 
       NGC1052 &      553300401 &   23.0 &   23.1 & (B) \\
               &      093630101 &   23.0 &   & \\
               &      553300301 &   23.0 &   & \\
               &      306230101 &   23.0 &   & \\
       NGC2273 &      140951001 &   23.2 &   24.0 & (B) \\
       NGC3227 &      101040301 &   22.9 &   22.8 & (A) \\
               &      400270101 &   22.2 &   22.0 & (T) \\
       NGC3516 &      107460601 &   22.5 &   & \\
               &      107460701 &   22.6 &   22.0 & (A) \\
               &      401210501 &   22.3 &   22.6 & (T) \\
               &      401210601 &   22.4 &   22.7 & (T) \\
               &      401210401 &   22.2 &   22.6 & (T) \\
               &      401211001 &   22.2 &   22.5 & (T) \\
       NGC4051 &      157560101 &   22.4 &   22.8 & (T) \\
               &      606321601 &   22.0 &   & \\
               &      606320201 &   22.1 &   & \\
               &      606321901 &   22.6 &   & \\
               &      606322001 &   22.4 &   & \\
               &      606320301 &   22.1 &   & \\
               &      606320401 &   22.6 &   & \\
               &      606321501 &   22.2 &   & \\
               &      606321701 &   22.3 &   & \\
               &      606322301 &   22.4 &   & \\
               &      606322201 &   22.5 &   & \\
               &      606321401 &   22.2 &   & \\
               &      606322101 &   22.7 &   & \\
               &      606321801 &   22.5 &   & \\
               &      606320101 &   22.3 &   & \\
       NGC4138 &      112551201 &   23.0 &   22.9 & (A) \\
       NGC4151 &      112310101 &   22.9 &   22.8 & (B) \\
               &      112830501 &   22.9 &   22.9 & (T) \\
               &      112830201 &   22.9 &   22.9 & (T) \\
               &      143500301 &   22.8 &   22.8 & (T) \\
               &      402660201 &   23.0 &   23.0 & (T) \\
               &      143500201 &   22.9 &   22.9 & (T) \\
               &      143500101 &   22.9 &   & \\
               &      402660101 &   23.1 &   & \\
       NGC4235 &      204650201 &   22.3 &   & \\
       NGC4258 &      400560301 &   22.9 &   22.8 & (A) \\
               &      110920101 &   23.0 &   22.9 & (C) \\
       NGC4388 &      110930701 &   23.5 &   23.6 & (B) \\
               &      675140101 &   23.4 &   & \\
       NGC4395 &      142830101 &   22.6 &   22.0 & (A) \\
               &      112521901 &   23.0 &   22.7 & (T) \\
       NGC5033 &      094360501 &   22.1 &   23.5 & (C) \\
       NGC5273 &      112551701 &   22.4 &   21.9 & (C) \\
       NGC5548 &      109960101 &   22.2 &   & \\
               &      089960301 &   22.1 &   & \\
               &      089960401 &   22.1 &   & \\ \hline
\end{tabular}
\caption{Logarithmic of the hydrogen column densities, $\rm{log(N_H)}$, for observations in our sample with $\rm{\nu_{SB}<10}$. References: (A) \citet{Akylas09}, (B) \citet{Brightman11}, (C) \citet{Cappi06}, and (T) \citet{Tombesi10}.}  
\label{tab:NHs}
\end{table}

\end{document}